\documentclass[%
 reprint,
superscriptaddress,
 amsmath,amssymb,
 aps,
]{revtex4-2}
\usepackage{siunitx}
\usepackage{graphicx}
\usepackage{bbm}
\usepackage{xcolor}
\definecolor{darkgreen}{rgb}{0.0,0.5,0.0}
\usepackage{amsfonts}
\usepackage{amstext}
\usepackage{array, booktabs}
\usepackage{longdivision}
\usepackage{float}
\usepackage{xr-hyper}     
\usepackage{hyperref}
\usepackage{soul}
\begin{document}
%\title{Atomic-scale origin of chiral domain walls in centrosymmetric FePd: beyond interface effects in the FePd/Nb hybrid system}
\title{Disorder-induced chirality in superconductor-ferromagnet heterostructures revealed by neutron scattering and multiscale modeling}
%\\
%\textcolor{blue}{Atomic disorder induces chiral magnetism in superconductor–ferromagnet heterostructures\\
%Disorder-driven chirality in superconductor–ferromagnet heterostructures revealed by neutron scattering and multiscale modeling}
%}

\newcommand{\UU}{Department of Physics and Astronomy, Uppsala University, Box 516, SE-75120 Uppsala, Sweden}

\newcommand{\LNU}{Department of Mathematics and Physics, Linnaeus University, SE-39231 Kalmar, Sweden}

\newcommand{\UFPA}{Faculdade de Física, Universidade Federal do Pará, CEP 66075-110, Belém, PA, Brazil}

\newcommand{\USP}{Instituto de Física, Universidade de São Paulo, CP 66318, 05315-970 São Paulo-SP, Brazil}

\newcommand{\ESS}{European Spallation Source ERIC, Partikelgatan 2, 224 84 Lund, Sweden}

\newcommand{\JUELICH}{Jülich Centre for Neutron Science for Quantum Materials and Collective Phenomena (JCNS-2), Forschungszentrum Jülich GmbH, 52425 Jülich, Germany}

\newcommand{\LUND}{Division of Synchrotron Radiation Research, Lund University, 22100 Lund, Sweden}

\newcommand{\JUELICHII}{Ernst Ruska-Centre (ER-C 2), Forschungszentrum Jülich GmbH, 52425 Jülich, Germany}

\newcommand{\ILL}{Institut Laue-Langevin, 71 Avenue des Martyrs, Grenoble Cedex 9 CS 20156, France}

\author{Annika Stellhorn}
\thanks{These two authors contributed equally.}
\affiliation{\ESS}

\author{Juan G. C. Palma}
\thanks{These two authors contributed equally.}
\affiliation{\USP}

\author{Alicia Backs}
\affiliation{\ESS}

\author{Anders Bergman}
\affiliation{\UU}

\author{Angela B. Klautau}
\affiliation{\UFPA}

\author{Emmanuel Kentzinger}
\affiliation{\JUELICH}

\author{Connie Bednarski-Meinke}
\affiliation{\JUELICH}

\author{Steffen Tober}
\affiliation{\JUELICH}

\author{Elizabeth Blackburn}
\affiliation{\LUND}

\author{Juri Barthel}
\affiliation{\JUELICHII}

\author{Nina-Juliane Steinke}
\affiliation{\ILL}

\author{Helena M. Petrilli}
\affiliation{\USP}

\author{Ivan P. Miranda}
\email{ivan.depaulamiranda@lnu.se}
\affiliation{\LNU}
\affiliation{\USP}

\date{\today}

\begin{abstract}
Chirality in superconductor-ferromagnet hybrids strongly influences phenomena such as the observable signatures of long-range triplet superconductivity, but its microscopic origin in nominally centrosymmetric ferromagnets is still unclear. Here, we combine structural characterization, polarization-analyzed grazing-incidence small-angle neutron scattering (PA-GISANS), first-principles calculations, and deep-learning-assisted multiscale modeling to study FePd and Nb/FePd heterostructures. Experimentally, we observe partial L1$_0$ order, atomic intermixing, anti-phase boundaries, and a depth-dependent defect gradient across the FePd layer, together with a finite net magnetic chirality at room temperature. The GISANS asymmetry indicates that the main chiral contribution lies in-plane, with an additional out-of-plane component associated with depth-dependent magnetic inhomogeneity. Theoretically, we show that chemical disorder in FePd, especially when combined with a compositional gradient, produces finite Dzyaloshinskii-Moriya interactions and stabilizes chiral finite-$\mathbf{q}$ magnetic modulations with mixed Bloch-Néel character. In the mesoscopic model, the resulting in-plane modulation length approaches the experimentally observed range. These results identify disorder and compositional gradients as intrinsic microscopic sources of net chirality in FePd-based films, showing that the observed chirality does not arise only from interface effects.
\end{abstract}

\maketitle

%\textcolor{blue}{Angela's notes in blue}
\vspace{10pt}

\section{Introduction}
% Now already included
%\textcolor{red}{Note Annika: Communications Materials is a journal for broad impact, focusing on whats currently missing in the field interesting for several communities. Shall we add at the end of the introduction an outlook for the usage of these methods towards a computation of chiralities in S/F structures as function of temperature or in general within the superconducting state, saying that our methods described here built the foundation for those studies? See note at the end of the Intro.}\\

It is known that superconductivity and ferromagnetism are antagonistic phenomena in nature: the former favors spin-singlet pairing, while the latter promotes spin polarization. Nevertheless, when a superconductor (S) is adjacent to a ferromagnet (F) to form a hybrid S/F system, a variety of nontrivial and technologically promising effects can emerge at the mesoscopic scale. These include a damped oscillatory behavior of the Cooper-pair wave function within the ferromagnetic medium \cite{Buzdin2005}, modifications of the superconducting critical temperature \cite{Linder2015}, the formation of so-called $\pi$ Josephson junctions with negative coupling \cite{Buzdin2005, Kulik1965}, and the confinement of superconductivity to magnetic domain walls \cite{Yang2004}, among others. Owing to these phenomena -- and their relevance for superconducting spintronics and quantum memory applications -- hybrid S/F systems continue to attract significant experimental and theoretical attention.\\

Within the broader class of S/F hybrid materials, \textit{chirality} plays a central role. Early theoretical works by Volkov \textit{et al.} \cite{Volkov2005,Volkov2003} predicted that a noncollinear alignment of magnetizations in S/F multilayer structures can generate long-range spin-triplet superconducting correlations, which were later observed experimentally \cite{DiBernardo2019}. These correlations give rise to an enhanced Josephson effect between superconductors separated by a ferromagnetic layer, with properties that depend sensitively on the real-space chirality of the magnetization texture \cite{Nikolic2025}. Moreover, chiral magnetic structures in S/F bilayers can significantly modify the superconducting critical temperature~\cite{Aladyshkin,Tagirov1999}, or even induce topological superconducting states that are robust against local disorder~\cite{Garnier, A.Svalland, Sardinero2024}. More recently attention has turned to S/F heterostructures hosting magnetic skyrmions in the F layer. In such systems, skyrmion-vortex pair formation has been shown both to host Majorana zero modes \cite{Nothhelfer2022,Rex2019} and to enable manipulation of the skyrmion size via proximity to the superconductor \cite{Dahir2019, S.S.Apostoloff, A.Petrovic, Xie2024}. Clearly, chiral S/F hybrid systems are at the intersection of two important research fields: chiral spintronics \cite{Yang2021} and superconducting spintronics \cite{Linder2015, Maggiora2024}.

Even though the mutual influence of chiral magnetic textures and superconducting states -- and their controllability via external magnetic fields -- has been actively investigated, comparatively little attention has been devoted to understanding the \textit{origin} of chirality in the ferromagnetic component of  S/F hybrid systems. Clarifying how chirality emerges in these materials, down to the microscopic level, is essential in developing realistic strategies to control and manipulate their mutual interaction.

% Magnetic chiral structures coupled with superconductivity induce exciting new quantum states technologically relevant for the field of spintronics, due to their controllability and colossal magnetoresistance effects \cite{SHYang, Linder2015}. 
% This new field of chiral spintronics holds great promise for the engineering of robust spin structures and topological states at the interface of the superconducting (S) and ferromagnetic (F) phases. 
% Furthermore, chiral structures in S/F bilayers are proposed to excite long-range supercurrents in S/F Josephson Junctions with chiral domain walls or other non-collinear arrangements of the ferromagnet \cite{Volkov2005, Robinson}, to impact on the superconducting critical temperature \cite{Aladyshkin, Tagirov1999}, or to induce a topological superconducting state stable against local disorder \cite{Garnier}. 
% The impact of magnetic chiral states on the superconducting state and their controllability by an external magnetic field is already well studied. 
% However, experimental observations of the reaction of the ferromagnetic state and its chiral structure are rare and yet poorly understood. 
 %It has been theoretically proposed that the chiral magnetic formation can be altered by the onset of superconductivity \cite{Buzdin2005}. 
 %A comprehensive understanding of this effect is essential for a successful application in spintronic devices.\\
 
Chirality manifests in the F layer through the formation of noncollinear spin textures, including skyrmions \cite{Dahir2019}, Bloch-type chiral domain walls (cDW), and Néel-type polarized structures~\cite{Yang2021}. In materials that are centrosymmetric in the bulk, such as FePd
%-- which has previously been studied in this context~\cite{Stellhorn2019} -- 
the emergence of chiral domain walls is commonly associated with the presence of a sizable Dzyaloshinskii-Moriya interaction (DMI)~\cite{Moriya,DMI_review}. Although DMI requires broken inversion symmetry to exist, such symmetry breaking can arise locally in nominally centrosymmetric systems due to structural or chemical inhomogeneities. 
%While the effects of DMI are well understood at the mesoscopic level, its microscopic origin at the atomistic scale remains far less explored. 
Recent studies have proposed several mechanisms capable of inducing local inversion-symmetry breaking, including strain effects~\cite{Allenspach2024}, the formation of antiphase boundaries~\cite{Chen2022}, and compositional intermixing~\cite{Pamela,Michels2019}, highlighting the importance of atomistic details in the emergence of chirality in such materials.
%A comprehensive understanding of this effect %\textcolor{red}{and its role on the hybrid S/F system} 
%is essential for a successful application in spintronic devices, especially because it has been theoretically proposed that the chiral magnetic formation can be altered by the onset of superconductivity \cite{Buzdin2005}. 

To address the origin of chirality in hybrid S/F materials, we investigate FePd-based heterostructures using a combined experimental and theoretical approach. Specifically, we consider relatively thick ($>40$ nm) L1$_0$-structured, nominally centrosymmetric FePd with perpendicular magnetic anisotropy (PMA), comparing a single FePd (F) film with a Nb/FePd (S/F) heterostructure. Although these are film samples, the FePd layer is grown on a thick Pd buffer and is correspondingly relaxed to a degree that, for the present purpose, allows its internal magnetic behavior to be treated as bulk-like. This makes it possible to investigate the inhomogeneity consequences of the chemical disorder and compositional gradient across FePd. Proximity effects associated with the Nb/FePd interface are not excluded, but they are not the primary focus of the present work; rather, our aim here is to establish whether the FePd layer itself already provides a microscopic route to chirality.
%Although interface proximity between FePd and Nb might play a role, here we focus on the inhomogeneity consequences of the chemical disorder and compositional gradient across FePd.
%Although interface proximity between FePd and Nb might play a role, it is to expect that such effects are confined around the interfacial region, and that, across FePd, it is superimposed by the bulk inhomogeneity consequences. 
%\textcolor{blue}{Perhaps we should be more cautious with this sentence. I don't know if, despite being local, Nb-FePd interdiffusion would influence, for example, the type of chirality (Neel X Bloch, etc.). I believe this doesn't invalidate the entire study, but maybe change it to something softer}. \textcolor{red}{Annika: right - not only interdiffusion, also sueprconducting effects can change the entire magnetic state of FePd. My suggestion: however, it is crucial to investigate the FePd bulk properties as a first step."}

To test this mechanism across the relevant length scales, we combine polarized grazing-incidence small-angle neutron scattering (GISANS), as well as macroscopic magnetic and structural measurements, with multiscale theoretical modeling. On the theory side, we employ a modified two-sublattice gradient model \cite{Gradient} together with first-principles density functional theory (DFT) calculations and a species-aware graph neural network (\texttt{SAGNN}) used as a surrogate for local magnetic moments and atomistic magnetic interactions in chemically disordered FePd. In this way, the machine-learning component is essential to bridge the atomistic DFT scale and the mesoscopic alloy models required to test whether disorder and defect gradients can account for the experimentally observed chirality.
%a combination of density functional theory and a 
%deep learning model based on graph neural networks (GNNs), here designed not only to fit this particular problem but also to fill the more general gap of obtaining realistic magnetic atomistic models for complex systems (e.g. alloys and amorphous) with an affordable computational cost. \\

We show that atomic disorder (intermixing), especially when associated with a compositional gradient across the sample, provides a microscopic mechanism for net chirality in the magnetic textures. This leads to a Bloch-Néel domain wall admixture that can be traced to atomistic DMI induced by the local breaking of inversion symmetry. Due to the strong Fe-Fe interactions, the dominant Bragg-like peaks of these chiral domain wall persist at room temperature. In particular, we show that disorder induces finite Fe-Fe DMI amplitudes comparable in magnitude to interface-driven DMI in ferromagnets/heavy-metal heterostructures such as Pd/Fe/Ir(111) \cite{Miranda2022} and Fe/Ir(111) \cite{Heinze2011}. The disorder-induced DMI is dominated by short-range pairs (within $r\lesssim1.5$ lattice parameters), and follows a statistically slower coherent decay rate with distance (more long-ranged) than the scaling $r^{-3}$ of RKKY-like interactions \cite{Pajda2001,Fert1980}. More broadly, this work establishes a microscopic and multiscale framework for investigating the origin and texture of chiral spin structures in disordered or partially ordered alloys, and provides a basis for future studies that explicitly incorporate superconducting effects in hybrid S/F systems.

First, macroscopic magnetic and structural characteristics of two PMA FePd heterostructures are investigated in Sections \ref{sec:structural}-\ref{sec:AFM-experimental}: (\textit{i}) the S/F heterostructure Nb/FePd (i.Nb/FePd), in comparison with (\textit{ii}) the single-film FePd (ii.FePd). Section \ref{sec:GISANS} explains in detail the observation of magnetic chirality in 
%with preferred handedness in the 
domain walls of FePd using polarization-analyzed GISANS (PA-GISANS) at room temperature, a tool that has already proven powerful for analyzing domain structures in ferromagnetic thin films \cite{Kentzinger2007}. Subsequently, the findings of structural defects and their possible impact on an average magnetic chiral spin structure are examined in Section \ref{sec:Theoretical} through a theoretical multiscale approach by first-principles calculations based on DFT, as well as atomistic and micromagnetic simulations, and \texttt{SAGNN} for a more efficient calculation of magnetic interactions. Taken together, the experimental and theoretical parts of this work address the same central question from complementary directions: the experiments establish the presence of structural disorder, depth-dependent inhomogeneity, and net magnetic chirality in FePd-based films, while the multiscale modeling tests whether such atomistic disorder is sufficient to generate finite DMI and preferred handedness. \\

%In summary, this paper presents a comprehensive study of the correlation between local defects and magnetic chirality in pristine magnetically isotropic ferromagnetic heterostructures through the formation of DMI, employing both experimental and theoretical approaches at the microscopic and macroscopic levels. This multiscale approach provides a foundation for integrating a range of experimental and theoretical methods. Our goal is to further investigate the observed correlations in the S/F system FePd/Nb, with the aim of gaining a deeper understanding of how magnetic chiral structures are entangled with the superconducting phase at the same time that we apply AI as an efficient tool for the analysis of magnetic interactions in complex systems.

%\textcolor{red}{maybe add?:}
%\begin{itemize}
    %\item kinds of chirality, what is been observed in FePd (using mostly CD-XRMS), 
    %\item current study: explain the shortly (in an overview) the experimental results and theoretical approach
    %\item polarized GISANS as tool to detect chirality (also not that common for itself!!!)
    %\item theoretical descriptions of such chiralities in pristine non-DMI systems also more or less new/not that common yet?
 %   \item combining all approaches and hence get a comprehensive overview
%\end{itemize}

\section{Experimental Results}\label{sec:Experimental-Results}

\subsection{Structural and magnetic properties of the L$1_0$ FePd films}\label{sec:structural} 

%%% 
%This section identifies possible structural defects that can be the origin of a non-zero DMI and evolution of spiral spin formations in components, which, in their pristine state, do not obtain any magnetic non-collinear configuration. 
This section identifies structural defects that may induce a finite DMI and thereby promote the emergence of noncollinear spin textures in a system that, in its pristine structure, would not be expected to host such magnetic configurations. One relevant example are the nominally centrosymmetric, L$1_0$-structured FePd films. In the perfectly ordered L$1_0$ phase, FePd consists of alternating Fe and Pd atomic layers, and crystallizes in a tetragonal lattice with $a=b\neq c$ (space group P4/\textit{mmm}), leading to a preferred magnetization along the $\hat{\mathbf{z}}=\hat{\mathbf{c}}=[001]$ direction~\cite{FePd4,Warren}. The long-range crystalline order of the L$1_0$-phase is defined via the order parameter $S$, ranging from a non-stochiometric sample and completely disordered Fe and Pd monolayers with $S=0$, to a fully long-range ordered state with $S=1$. Most importantly, the structural long-range order impacts on the magnitude of the PMA through spin-orbit coupling: higher $S$ leads to higher PMA \cite{Gehanno1999}.\\

For the growth of high-quality L$1_0$-FePd films, we employed Molecular Beam Epitaxy (MBE) as described in detail in the Methods, Section \ref{sec:Suppl-Experimental}. Two PMA samples with large FePd thickness ($>40$ nm), both adjacent to thick Pd  layers to adjust the lattice constant from MgO to FePd, i.Nb/FePd and  ii.FePd, are compared; the samples specifications are also described in the Methods, Section \ref{sec:Suppl-Experimental}. Further, Section \ref{sec:Suppl-Experimental} also describes in detail how to calculate the uniaxial anisotropy constant $K_u$ from bulk magnetization measurements via SQUID magnetometry (MPMS), and the structural long-range order parameter $S$ of the FePd L1$_0$-phase from the ratio between the FePd(001) superlattice and FePd(002) fundamental reflections measured by X-ray Diffraction (XRD)  \cite{DoktorarbeitGehanno}.

\begin{figure}[t!]
    \centering
    \includegraphics[width=1\linewidth]{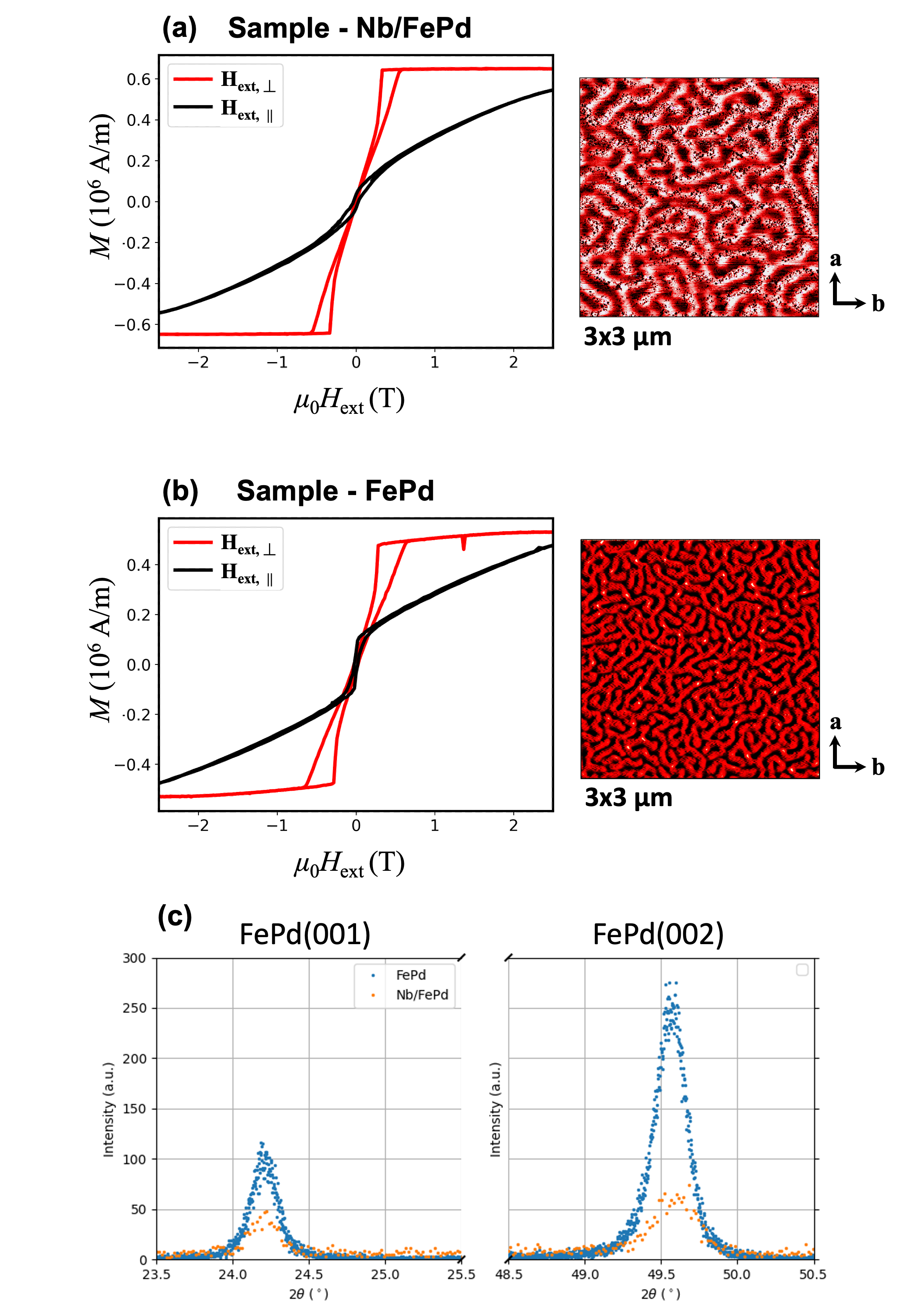}
    \caption{\textbf{Magnetic structure of i.Nb/FePd and ii.FePd.} (a,b) Hysteresis loops obtained by MPMS, and surface magnetic domain structure obtained by MFM of i.Nb/FePd (top) and ii.FePd (bottom). (c) XRD measurements of the FePd(002) and FePd(001) Bragg peaks, with the peak ratio indicating the quality of the L$1_0$ phase of FePd. All measurements are obtained at 300~K.}
    \label{fig:MPMS-MFM}
\end{figure}

Table \ref{tab:Ku-S-parameters} summarizes $K_u$ and $S$ for both systems, obtained from the bulk magnetization and XRD measurements shown in Fig. \ref{fig:MPMS-MFM}(c). The films exhibit partial long-range structural order ($0<S<1$), indicating deviations from the perfectly ordered tetragonal L$1_0$ structure, in which
%meaning that they partially consist of FePd in the tetragonally ordered P4/\textit{mmm} crystalline structure, and partially in the disordered phase with intermixed Fe and Pd atoms. 
i.Nb/FePd shows a higher $S$ value compared to ii.FePd. An incomplete degree of structural order can be caused by structural defects such as intermixing, planar defects including anti-phase boundaries, and a compositional gradient along the growth direction, which will be characterized further below. 
%Despite the relatively strong spin-orbit coupling in FePd causing the uniaxial magnetocrystalline anisotropy in the tetragonally distorted L1$_0$-phase, 
Moreover, a partial long-range order is connected to a reduced saturation magnetization and reduced $K_u$. Magnetization measurements reveal a strong uniaxial anisotropy in both systems, but a higher $K_u$ in i.Nb/FePd compared to ii.FePd, consistent with the obtained $S$ values. As expected for the PMA in both systems, Fig. \ref{fig:MPMS-MFM}(a,b) reveals a maze-like magnetic domain pattern along the $\pm\hat{\mathbf{z}}$ directions in both samples.\\

%The measured structural and magnetic information of both samples is summarized in Table \ref{tab:Ku-S-parameters}. i.Nb/FePd obtains a higher degree of long-range structural order than sample ii.FePd. The imperfect long-range order of both samples is due to structural defects such as intermixed Fe and Pd atomic positions, interdiffusion, or planar defects like anti-phase boundaries. 

\begin{table}[h]
    \centering
    \caption{\textbf{Structural long-range order parameter $S$ and uniaxial anisotropy constant $K_u$ of i.Nb/FePd and ii.FePd.}}
    \begin{tabular}{@{}lcc@{}}
        \toprule
        \textbf{Parameter} & \textbf{i.Nb/FePd} & \textbf{ii.FePd} \\
        \midrule
        $S$ & $0.70 \pm 0.01$ & $0.52 \pm 0.01$ \\
        $K_u$ (MJ/m$^3$) & $1.5 \pm 0.2$ & $1.0 \pm 0.2$ \\
        \bottomrule
    \end{tabular}
    \label{tab:Ku-S-parameters}
\end{table}

%Possible defect sites leading to partial long-range order ($S<1$), and non-zero average DMI values occur due to: (i) a random intermixing of Fe and Pd atoms and a gradient defect distribution, and (ii) anti-phase-boundaries. \\

\subsection{Intermixing and compositional gradient distribution}\label{sec:TEM-experimental} 
%\textcolor{red}{add new description for explanation S from peakratio, and more adapted to figure}

%\onecolumn
\begin{figure}[H]
    \centering
    \includegraphics[width=1.1\linewidth]{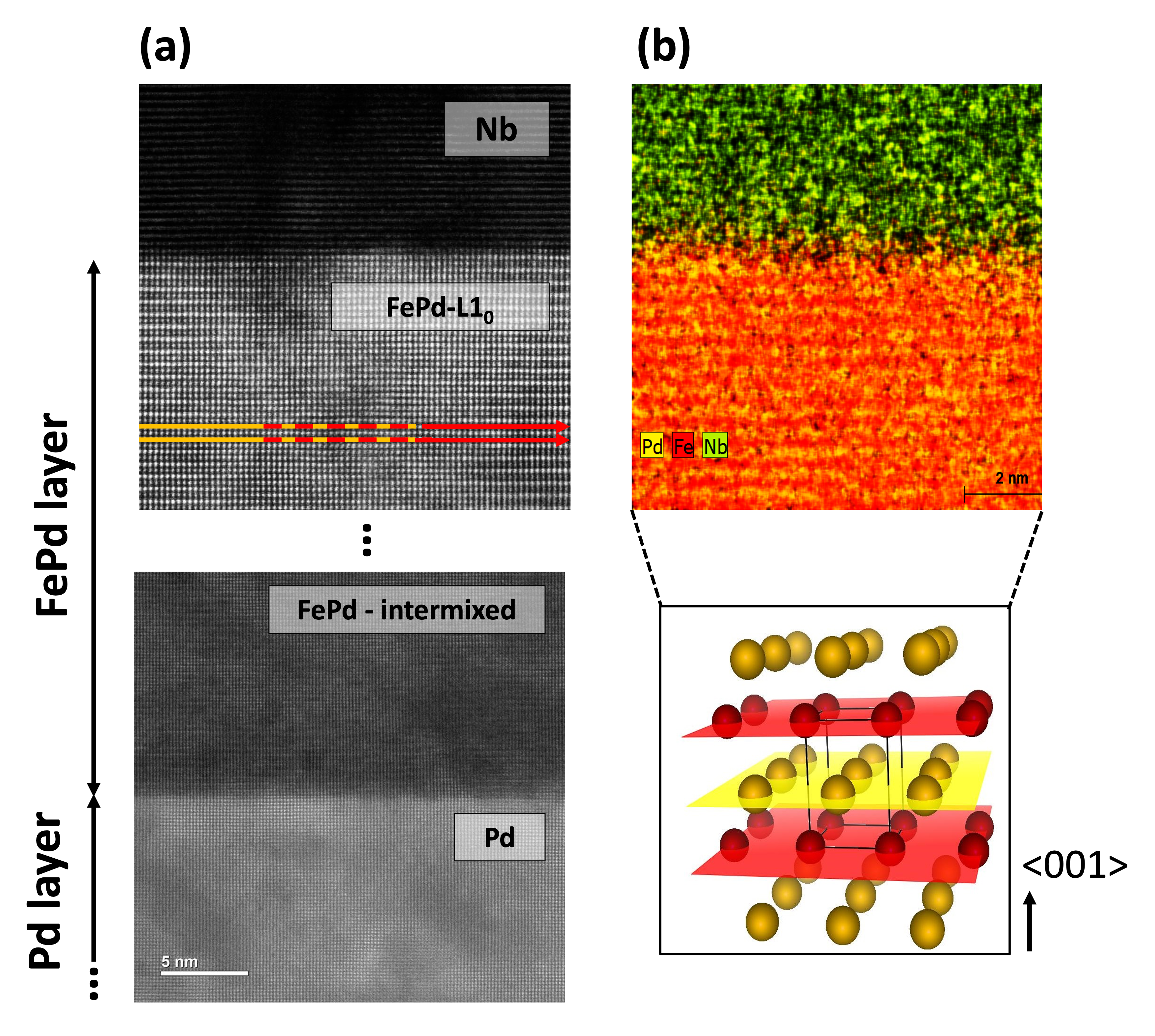}
    \caption{\textbf{Structural characterization obtained at 300~K}. (a) HAADF-STEM measurements of the Nb/FePd interface in i.Nb/FePd (top) and the FePd/Pd interface (bottom). (b) HAADF-STEM EDX measurements of the Nb/FePd interface in i.Nb/FePd (top), and the respective schematic monolayer structure of the L$1_0$ phase of FePd with alternating Fe (red) and Pd (yellow) layers (bottom). 
    %Planar defects inside Pd result in a defect-rich FePd layer with lower defect-density towards the Nb/FePd surface, a so-called "defect-density gradient". Inset in (b) schematically shows the perfect L$1_0$-structured FePd ordering, mostly visible near the Nb/FePd interface. Images are adapted from \cite{PhDStellhorn}.
    }
    \label{fig:XRD-HR-TEM}
\end{figure}
%\twocolumn

%For the High-Angle Annular Dark-Field Scanning Transmission Electron Microscopy (HAADF-STEM) 
%\textcolor{green}{Juan: Is it okay to first mention this measurement with the acronyms directly? \textcolor{red}{You are right, better not}} measurements, we have used a FEI Titan G2 80-200 ChemiSTEM at ER-C, research center Juelich \cite{ERC-Juelich}. XRD measurements have been performed using a Bruker AXS D8 Advanced system, AFM/MFM measurements have been done in the ac intermittent tapping mode of an Agilent 5400 AFM/SPM probe with HQ:NSC36/Co-Cr/Al type multi-cantilevers, \textcolor{red}{and bulk magnetization for magnetic hysteresis loops have been measured using a Quantum Design Magnetic Properties Measurement System (MPMS) based on a rf-SQUID.}\\
The partial long-range order identified in Section \ref{sec:structural} already implies a local symmetry breaking in FePd. Among the resulting defects, intermixing and depth-dependent compositional gradients are particularly relevant here. While both samples exhibit such disorder, i.Nb/FePd is used below as the most illustrative example in the microscopy analysis. 

Figure \ref{fig:XRD-HR-TEM}(a,b) shows high-resolution HAADF-STEM and HAADF-STEM EDX elemental maps of i.Nb/FePd, respectively. In a well-ordered L$1_0$ FePd structure, the alternating Fe and Pd layers produce a clear contrast in the STEM image due to their different scattering cross sections, whereas defect-richer regions appear with much weaker contrast. 
%In Fig. \ref{fig:XRD-HR-TEM}(a), the bottom Pd layer contains several planar defects, marked by yellow arrows, which can propagate across the Pd/FePd interface and contribute to the formation of a defect-richer FePd region.

This becomes evident in the reduced contrast between atomic rows near the FePd/Pd interface (middle image). Such contrast loss may arise either from Fe-Pd intermixing or from the projection of multiple planar defects within the L$1_0$-structure FePd layer. One planar defect that converts an Fe monolayer into a Pd monolayer along the horizontal direction is highlighted by the yellow-to-red guide lines in the top image of Fig. \ref{fig:XRD-HR-TEM}(a). In addition, electron microscopy reveals a clear defect-density gradient: the FePd region shows less contrast, possibly signaling intermixed Fe and Pd atoms, near the Pd/FePd interface (lower image) and becomes progressively more ordered towards the Nb/FePd interface (upper image), where the L$1_0$ layering is much cleaner and fewer lattice defects remain. Consistently, the EDX elemental map in Fig. \ref{fig:XRD-HR-TEM}(b) shows a well-defined alternation of Fe and Pd planes close to the Nb/FePd interface.

%Whereas a monolayer structure is visible via different scattering cross sections and a clear contrast between Fe and Pd atoms, a defect-rich state would lead to a lower contrast. 
%The bottom Pd layer in Fig. \ref{fig:XRD-HR-TEM}(a) contains several planar defects indicated by yellow arrows, which at the FePd/Pd interface can lead to a defect-rich FePd layer. In the HAADF-STEM image at the FePd/Pd interface (middle image), this becomes obvious through a very low contrast between the atom rows. This can result either from an overlay of planar defects within an L$1_0$-structured FePd layer along the projection direction, or from an intermixing of Fe and Pd atoms. One planar defect leading to a transition of the Fe to a Pd monolayer along the horizontal direction is marked by yellow-to-red planar lines in Fig \ref{fig:XRD-HR-TEM}(a, top). In addition, a gradient within the defect density from the FePd/Pd interface at the bottom towards the Nb/FePd interface at the top is visible, compared to a much cleaner L$1_0$-phase and only few lattice defects near the Nb/FePd interface. The EDX elemental map in Fig. \ref{fig:XRD-HR-TEM}(b) indicates the formation of a clear monolayer structure with alternating Fe and Pd planes near the Nb/FePd interface.\\

The structural defect gradient along $\hat{\mathbf{z}}$ can induce, due to spin-orbit coupling, a depth-dependency of the local DMI and anisotropies. This may favor chiral magnetic textures with a finite out-of-plane component. 
%a magnetic gradient, leading to magnetic chiral structures with propagation vector components parallel to $z$. 
These signatures, associated with the $\hat{\mathbf{c}}\parallel\hat{\mathbf{z}}$ direction, 
%These chiral propagations with chiral vector $\mathbf{c}||\mathbf{z}$ 
will be discussed in Section \ref{sec:GISANS}. A further possible contribution to such depth-dependent magnetic inhomogeneity is strain along $\hat{\mathbf{z}}$, related to the film growth conditions.
%Another possible reason for a structural defect gradient, causing a magnetic gradient along the depth of FePd, could be strain along $<$001$>$ due to the thin film growth conditions. 
However, because both the Pd buffer and the FePd layer are relatively thick, and the lattice mismatch is small ($\sim0.3\%$) \cite{DoktorarbeitGehanno}, long-range coherent epitaxial strain is expected to be largely relaxed and is therefore unlikely to be the dominant origin of the observed gradient. By contrast, defect formation and chemical disorder provide a more natural explanation for the depth-dependent inhomogeneity identified here, although local strain effects cannot be completely excluded.
%However, both the Pd buffer layer as also the FePd layer are thick films $>$ 40~nm, with a lattice mismatch of about 0.3$\%$ \cite{DoktorarbeitGehanno}. Defect formations within the FePd or Pd layers are rather likely to form than epitaxial strain due to lattice mismatch.

\subsection{Anti-phase boundaries}\label{sec:AFM-experimental}

\begin{figure}[H]
    \centering
    \includegraphics[width=1\linewidth]{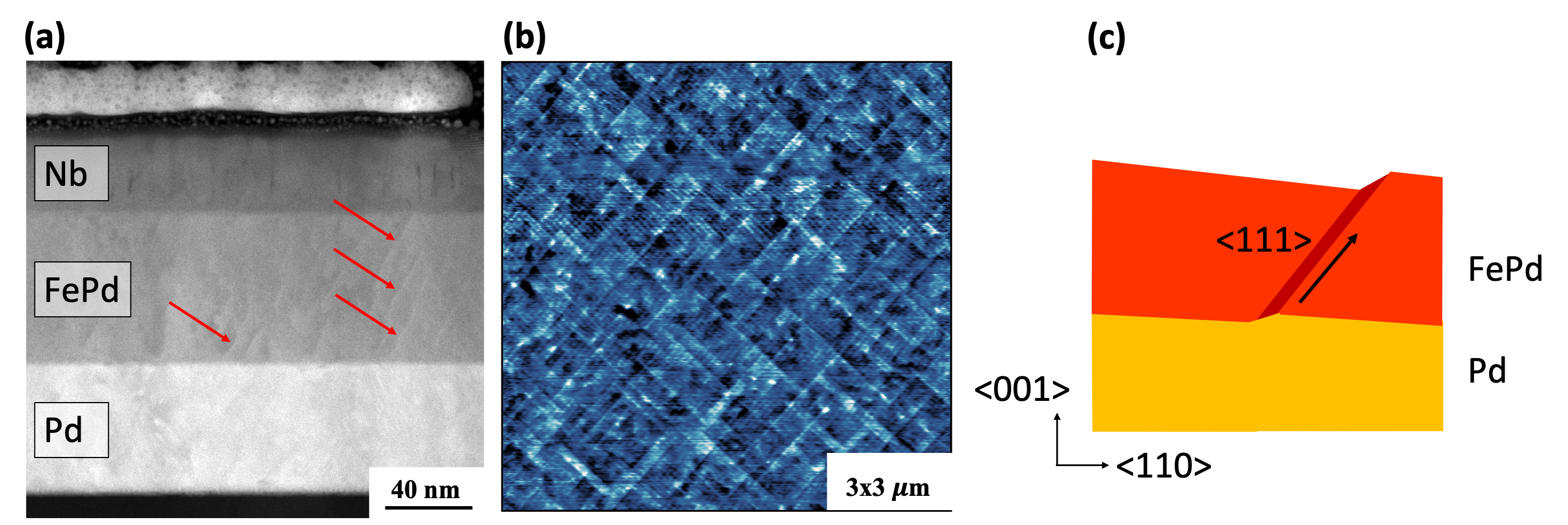}
    \caption{\textbf{Anti-phase boundaries in i.Nb/FePd}. (a) Cross section of the sample obtained from HAADF-STEM measurements, (b) AFM image showing the top surface morphology, and (c) schematic cross-section of anti-phase boundaries along $[111]$. All measurements were performed at 300~K. Images are adapted from \cite{PhDStellhorn}.}
    \label{fig:XRD-HR-TEM2}
\end{figure}

 Anti-phase boundaries have been observed in both i.Nb/FePd and ii.FePd samples using HAADF-STEM, XRD and AFM: planar defects (anti-phase boundaries) along the (111) plane are marked by red arrows in Fig. \ref{fig:XRD-HR-TEM}(a). These anti-phase boundaries break up into facets
 %terraces 
 along $[110]$ at the surface layer, as shown in the AFM image of Fig. \ref{fig:XRD-HR-TEM}(b). A schematic representation of this defect geometry is given in Fig.~\ref{fig:XRD-HR-TEM}(c). Such anti-phase boundaries provide an additional source of structural inhomogeneity in the FePd layer. By locally disturbing the ideal L$1_0$ structure, they contribute to changes in the local magnetic interactions and possibly to the emergence of domain walls \cite{Nieves2017} and chiral magnetic textures. However, even though in the present work they are identified as a relevant structural defect class, the microscopic modeling focuses on the effects of chemical intermixing and compositional gradients discussed above.
 
 %In the present work, they should be viewed as a plausible defect class acting alongside intermixing and the depth-dependent defect gradient discussed above.

%%%% The text below was moved to the next section.
%In summary, Sections \ref{sec:structural}-\ref{sec:AFM-experimental} highlight the origins of defect structures in the  FePd layer in the heterostructures. In FePd thin films, the magnetic structure is tightly bound to the structural long-range order of the L$1_0$-phase. Intermixed Fe and Pd sites, a gradient defect distribution, and anti-phase boundaries can lead to non-zero DMI and spiral spin structures. Two high-PMA samples have been investigated, where sample i.Nb/FePd has a higher degree of long-range structural order than ii.FePd, indicated as well by a lower saturation magnetization and lower $K_u$ as compared to FePd. In the next sections, the effect on a formation of magnetic chiral structures will be first described experimentally (section \ref{sec:GISANS}) via PA-GISANS, and second from a theoretical approach based on first-principles as well as atomistic, micromagnetic simulations (section \ref{sec:Theoretical}).\\  

%\textcolor{red}{maybe Ku and S can be correlated to more chirality - 622: $Ku = 1.5\pm0.2\cdot10^6\,$J/m$^3$), $S = 0.70\pm0.01$, 505: $Ku = 1.0\pm0.2\cdot10^6\,$J/m$^3$, $S = 0.52\pm0.01$ - but are two samples enough to say that?? I only have results from these two from GISANS}\\

\subsection{Determination of chirality via PA-GISANS}\label{sec:GISANS}

Taken together, Sections \ref{sec:structural}-\ref{sec:AFM-experimental} %highlight the origins of defect structures in the  FePd layer in the heterostructures.
show that the FePd layers are not perfectly ordered L$1_0$ systems, but contain several forms of structural inhomogeneity, including intermixing, a depth-dependent defect gradient, and anti-phase boundaries.

In PMA systems with stripe- or maze-domain patterns, chirality is most naturally manifested in the magnetic domain walls separating oppositely magnetized domains. Their lateral periodicity and depth dependence make them well suited for investigation by PA-GISANS; the experimental details are given in the Methods, Section \ref{sec:Suppl-Experimental}. This technique probes nuclear and magnetic lateral correlations on nanometer and mesoscopic length scales in a depth-sensitive geometry. It is performed where the incident angle $\theta_i$ is close to the angle of total reflection $\theta_C$, in order to maximize the scattering cross section while still probing the full layer thickness \cite{Kentzinger2008}. Within the small-angle approximation, the components of the scattering vector $\mathbf{Q}$ are determined by $\theta_i$ and by the scattering angles $\Delta\theta_x$ and $\Delta\theta_y$, defined in the geometry shown in Fig.~\ref{fig:2D-GISANS} of Section~\ref{sec:methods} (see also Ref.~\cite{Fermon1999}):

\begin{eqnarray}
    \begin{pmatrix}
    Q_x \\ Q_y \\ Q_z\\ 
    \end{pmatrix}
    = \frac{2\pi}{\lambda_n}
    \begin{pmatrix}
    \theta_i\Delta\theta_x + \frac{(\Delta\theta_x)^2}{2} + \frac{(\Delta\theta_y)^2}{2}\\ 
    \Delta\theta_y\\
    2\theta_i+\Delta\theta_x\\
    \end{pmatrix},
\end{eqnarray}

\noindent where $\lambda_n$ is the neutron wavelength. Using uniaxial polarization analysis, the direction, chirality, and nuclear-magnetic interactions of magnetic moments can be determined for specific polarization directions \cite{Tasset1999, Schweika2010}. Here, we concentrate on the analysis of chiral spin structures through the spin-asymmetry within the GISANS scattering peaks, i.e., the difference of spin-flip (SF) cross sections along a given polarization axis $\boldsymbol{\nu}$, which is given by:

\begin{equation}
\Delta I_{\boldsymbol{\nu}}^{SF}
=
\frac{1}{2}(I^{+-}-I^{-+})_{\boldsymbol{\nu}}
=
2i\mathbf{C}(\mathbf{Q})\cdot\boldsymbol{\nu}
=
2iC_{\boldsymbol{\nu}}(\mathbf{Q})
\label{eq:asymmetry}
\end{equation}

\noindent where

\begin{equation}
\label{eq:asymmetry2}
\mathbf{C}(\mathbf{Q})=\left[\mathbf{M}_{\perp}^*(\mathbf{Q})\times\mathbf{M}_{\perp}(\mathbf{Q})\right]_{\boldsymbol{\nu}}.
\end{equation}

\begin{figure}[H]
    \centering
    \includegraphics[width=1\linewidth]{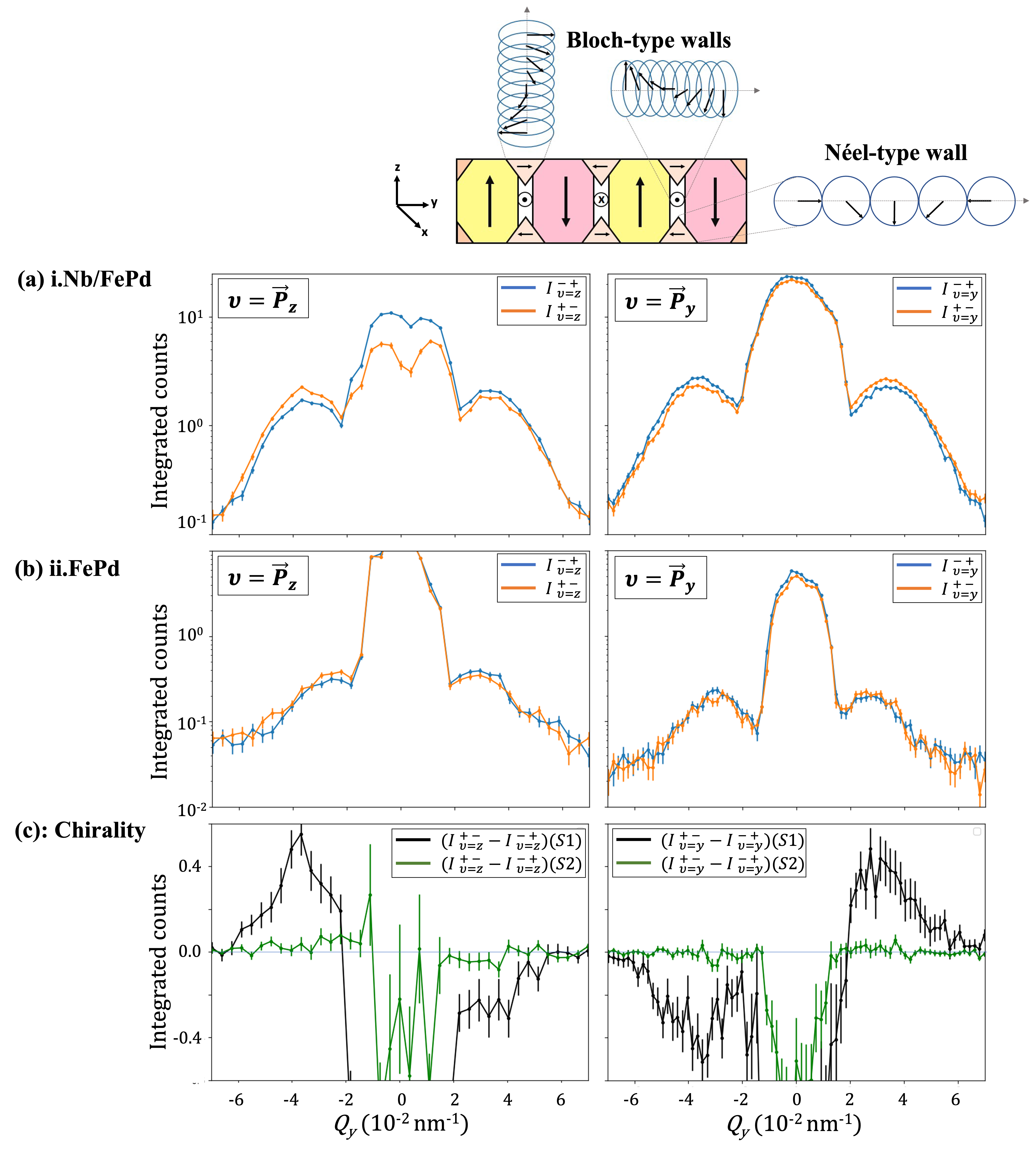}
    \caption{\textbf{PA-GISANS measurements and extracted chirality.} $I(Q_y)$ plots of (a) i.Nb/FePd and (b) ii.FePd, in comparison with (c) the respective chirality plots $(I^{+-}-I^{-+})(Q_y)$. Displayed for a polarization along the surface normal $<$001$>$ (left), and within the surface plane (100) (right). All measurements were made at 300~K. }
    \label{fig:polGISANS-together-SF}
\end{figure}

In Eqs. \ref{eq:asymmetry}-\ref{eq:asymmetry2}, $I^{+-}_{\boldsymbol{\nu}}$ and $I^{-+}_{\boldsymbol{\nu}}$
denote the spin-flip scattering intensities for incident and outgoing neutron spin states defined with respect to $\boldsymbol{\nu}$. The quantity $\mathbf{M}_{\perp}(\mathbf{Q})$ is the component of the magnetic scattering amplitude perpendicular to the scattering vector $\mathbf{Q}$, i.e.
$\mathbf{M}_{\perp}=\hat{\mathbf{e}}_{\mathbf{Q}}\times(\mathbf{M}\times\hat{\mathbf{e}}_{\mathbf{Q}})$. Eq.~\ref{eq:asymmetry} shows that the spin-flip asymmetry directly probes the chiral contribution to the magnetic scattering through the projection of $\mathbf{C}$ onto the polarization axis. Thus, the asymmetry vanishes unless the magnetic texture has a finite chiral component along $\boldsymbol{\nu}$.
%According to Eq. \ref{eq:asymmetry}, the asymmetry in the GISANS peak intensities between the spin-flip channels "+-" and "-+", is directly connected to the magnetic chirality $\mathbf{C}$ and can be measured only with a polarization direction $\boldsymbol{\nu}$ along the direction of chirality $\mathbf{C}$. 
As sketched in the magnetic model of the ferromagnetic FePd layer cross-section on top of Fig. \ref{fig:polGISANS-together-SF}, chiral domain walls in FePd films with high PMA are expected to exist mainly as Bloch-type domain walls separating the out-of-plane oriented domains \cite{Hubert2008}. In films with lower PMA, chiral closure domains of Néel-type can also exist, seen schematically as triangular shaped surface closure domains at both interfaces. 
%but due to the very high $K_u$-values of i.Nb/FePd and ii.FePd, these are not taken into account here. 
In both cases, the associated chiral contributions are projected mainly onto the in-plane directions $\hat{\mathbf{x}}$ and $\hat{\mathbf{y}}$ of the film. Because the FePd samples studied here exhibit a maze-like domain pattern (see Fig.~\ref{fig:MPMS-MFM}(b)), the experiment does not resolve contributions along $\hat{\mathbf{x}}$ and $\hat{\mathbf{y}}$ separately. We therefore probe only one representative in-plane direction, $\hat{\mathbf{y}}$, following the usual GISANS notation. Consequently, the measured scattering asymmetry reflects only a net chirality, i.e. an overall preferred handedness, together with a net in-plane domain-wall orientation.

Two PA-GISANS geometries were employed at room temperature, as described in the Methods Section~\ref{sec:methods}. In the first geometry, the magnetic guide field $\mathbf{B}$ was applied perpendicular to the sample surface, i.e., along the surface normal ($[001]$, parallel to the $a$-$b$-plane). In the second geometry, $\mathbf{B}$ was applied parallel to the sample surface, i.e., within the $ab$ plane and along the in-plane direction associated with ${Q}_y$. The corresponding GISANS data, measured at the total-reflection edge, are shown in Fig.~\ref{fig:polGISANS-together-SF}(a,b). As mentioned, the schematic above Fig.~\ref{fig:polGISANS-together-SF} indicates the possible directions of magnetic chirality associated with Néel- and Bloch-type domain walls in L1$_0$-ordered FePd. For both geometries, the GISANS peaks of i.Nb/FePd and ii.FePd appear at the angular and reciprocal-space positions listed in Table~\ref{tab:Q-angles}. In the following, we assume scattering from the GISANS line with $\Delta\theta_x=0$ and $\theta_i=\theta_C$.
%Two different sample geometries were employed in PA-GISANS at room-temperature as described in the Methods Section~\ref{sec:methods}: with the applied magnetic guide field $\mathbf{B}$ (a) perpendicular to the sample surface (i.e., perpendicular to $<$001$>||\,\mathbf{Q}_z$, and parallel to the $a$-$b$-plane), and (b) parallel to the sample surface (i.e., parallel to the plane $a$-$b$-plane || (100)$\,||\,\mathbf{Q}_y$). Fig. \ref{fig:polGISANS-together-SF}(a-b) show GISANS measurements at the total reflection edge. The sketch on top of Fig. \ref{fig:polGISANS-together-SF} denotes possible directions of magnetic chiralities in Néel- or Bloch-type domain walls within L1$_0$-ordered FePd. 
%For both geometries, GISANS peaks of i.Nb/FePd and ii.FePd appear at the angular- and Q-values given in table \ref{tab:Q-angles}. We assume scattering purely from the GISANS line, with $d\theta_x\,=\,0$ and $\theta_i\,=\,\theta_{C}$. 

\begin{table}[htb!]
    \centering
    \caption{\textbf{GISANS peak characteristics of i.Nb/FePd and ii.FePd}.}
    \begin{tabular}{@{}lcc@{}}
        \toprule
        \textbf{Value} & \textbf{i.Nb/FePd} & \textbf{ii.FePd} \\
        \midrule
        $\theta_y$ (deg) & $0.25 \pm 0.02$ & $0.21 \pm 0.03$ \\
        $\theta_i$ (deg) & $0.558 \pm 0.008$ & $0.62 \pm 0.08$ \\
        $Q_x$ ($\mathrm{nm}^{-1}$) & $(4.8\pm0.6)\times10^{-4}$ & $(3.5\pm0.8)\times10^{-4}$ \\
        $Q_y$ ($\mathrm{nm}^{-1}$) & $0.034 \pm 0.002$ & $0.029 \pm 0.003$ \\
        $Q_z$ ($\mathrm{nm}^{-1}$) & $0.153 \pm 0.02$ & $0.17 \pm 0.02$ \\
        \bottomrule
    \end{tabular}
    \label{tab:Q-angles}
\end{table}

We analyzed the GISANS peak asymmetry, $\Delta I_{\boldsymbol{\nu}}^{SF}$, for the two measurement geometries shown in Fig.~\ref{fig:polGISANS-together-SF}: left column, with polarization axis out of plane ($\boldsymbol{\nu}=\mathbf{P}_z$), and right column, with polarization axis in plane ($\boldsymbol{\nu}=\mathbf{P}_y$). The corresponding averaged spin-flip intensity differences are displayed in Fig.~\ref{fig:polGISANS-together-SF}(c), and the integrated values are summarized in Table~\ref{tab:Asymmetry}.

%We have analyzed the GISANS peak asymmetry $\Delta I_{\boldsymbol{\nu}}^{SF}$ for contributions of $\mathbf{C}$ for case (a) with polarization axis out-of-plane ($\boldsymbol{\nu}=\mathbf{P_z}$) and case (b) with polarization axis in-plane ($\boldsymbol{\nu}=\mathbf{P_y}$) with respect to the thin film surfaces (see insets in Fig. \ref{fig:polGISANS-together-SF}). The averaged difference of intensities of the SF spin channels is displayed in Fig. \ref{fig:polGISANS-together-SF}(c), with integrated scattering intensities listed in table \ref{tab:Asymmetry}

%Values for an integration between $Q_y=\pm[0.002,0.0065]\,$nm$^{-1}$ (to improve statistics and to exclude the specular peak contribution), are listed in table \ref{tab:Asymmetry}. \\

%i.Nb/FePd Pz = 2.5
%ii.FePd Pz = 0.3
%i.Nb/FePd Py = 5.8
%ii.FePd Py = 0.2
\begin{table}[htb!]
    \centering
    \caption{\textbf{Asymmetry of GISANS peaks}, i.e., $\frac{1}{2}(I^{+-}-I^{-+})$, integrated over $\mathbf{Q_y}=\pm[0.002,0.0065]\,$nm$^{-1}$ and $\mathbf{Q_z}=[0.0137,0.02]\,$nm$^{-1}$, and averaged between left and right GISANS peaks. The values represent a relative comparison between chiral structures along $\hat{\mathbf{y}}$ and $\hat{\mathbf{z}}$ for i.Nb/FePd and ii.FePd.}
    \begin{tabular}{@{}lcc@{}}
        \toprule
        \textbf{Geometry} & \textbf{i.Nb/FePd (counts)} & \textbf{ii.FePd (counts)} \\
        \midrule
        $\boldsymbol{\nu}=\mathbf{P_z}\,||\,[001]$ & $2.5 \pm 0.2$ & $0.3 \pm 0.1$ \\
        $\boldsymbol{\nu}=\mathbf{P_y}\,||\,(100)$ & $5.8 \pm 0.3$ & $0.2 \pm 0.1$ \\
        \bottomrule
    \end{tabular}
    \label{tab:Asymmetry}
\end{table}
%see script /Users/annikastellhorn/Desktop/OneDrive-Vorher/ESS Rechner/Neutron proposals and reports/ILL submission Sep 2022/D33 experiment/Analysis/scripts Alex new script
%( Asymm (diffference) rechts + Asymm (difference) links )/2

Sample i.Nb/FePd exhibits a larger spin-flip asymmetry than ii.FePd for both polarization directions, indicating a stronger net chiral contribution to the magnetic scattering and, therefore, a more pronounced preferred handedness of the domain-wall texture. The difference between both samples indicates that the net chiral response is sensitive to the underlying structural and magnetic state of the FePd layer. Although i.Nb/FePd exhibits both larger asymmetry and higher values of $S$ and $K_u$ than ii.FePd, the present comparison does not establish a direct monotonic relation between chirality and either parameter separately. Rather, it supports the interpretation that the observed chiral response depends on the overall defect landscape, including not only the degree of long-range order, but also the type and spatial distribution of structural inhomogeneities.

For i.Nb/FePd, the asymmetry is markedly larger for $\boldsymbol{\nu}=\mathbf{P}_y$ than for $\boldsymbol{\nu}=\mathbf{P}_z$, indicating that the dominant chiral contribution is associated with in-plane components of the texture, as expected for Bloch- and Néel-type domain walls in PMA systems. At the same time, the finite asymmetry observed for $\boldsymbol{\nu}=\mathbf{P}_z$ points to a non-vanishing out-of-plane projected chiral contribution. A natural interpretation is that this contribution is linked to the depth-dependent structural inhomogeneity identified in Section~\ref{sec:TEM-experimental}, which can induce a corresponding depth dependence of the local magnetic interactions.

%i.Nb/FePd has a higher SF asymmetry than ii.FePd, indicating a larger net chiral contribution to the magnetic scattering. This is consistent with a stronger preferred handedness of the domain wall texture. The difference between both samples may be caused by the different defect structure of both systems, which is also seen in their differing $S$ and $K_u$ values: ii.FePd has a lower long-range structural order leading to a lower magnetocrystalline anisotropy constant. This is due not only to an increased amount of defect-sites, but also to other kinds of structural defects from the slightly different growth parameters (e.g., evaporation rate and resulting film thickness) of each sample. As expected for the evolution of Bloch and Néel domain walls in high-PMA FePd along the in-plane direction, the net-chirality of i.Nb/FePd in the in-plane direction along $\mathbf{y}$ is larger than in the out-of-plane direction along $\hat{\mathbf{z}}$. In summary, i.Nb/FePd shows clear chiral signals with a propagation vector along both the in-plane $\mathbf{y}$ and out-of-plane $\hat{\mathbf{z}}$ directions. 
%Typically, both Bloch and Néel domain walls propagate along the in-plane direction, whereas an out-of-plane chirality can be caused by the structural gradient diffusion along $\hat{\mathbf{z}}$ that has been shown in Section \ref{sec:TEM-experimental}, which, via the spin-orbit coupling in FePd, leads to magnetic inhomogeneity along $\hat{\mathbf{z}}$.

%thickness SN505: 55\pm2, SN622:44\pm5

In the following, we present a theoretical multiscale approach combining first-principles DFT calculations, atomistic simulations, and deep-learning-assisted modeling to understand how different defect types can generate distinct chiral magnetic structures through the emergence of DMI in FePd film heterostructures, and to compare these results with the experimental observations.

%In the following sections, we describe a theoretical multiscale approach by first-principles calculations based on DFT, as well as atomistic and deep-learning-based simulations aiming to understand how different defect types can lead to distinct chiral magnetic structures via the evolution of DMI in thin film FePd heterostructures, and compare the results with the previously shown measurements. \textcolor{red}{Note Annika: check at the very end the overall comparison between experimental and theoretical results!} 
%\textcolor{red}{Qz not really interesting here}\\

\section{Theoretical Results}\label{sec:Theoretical}

\subsection{Ordered L$1_0$ FePd: electronic structure and magnetic properties}

\begin{figure}[h]
    \centering
    \includegraphics[width=0.9\linewidth]{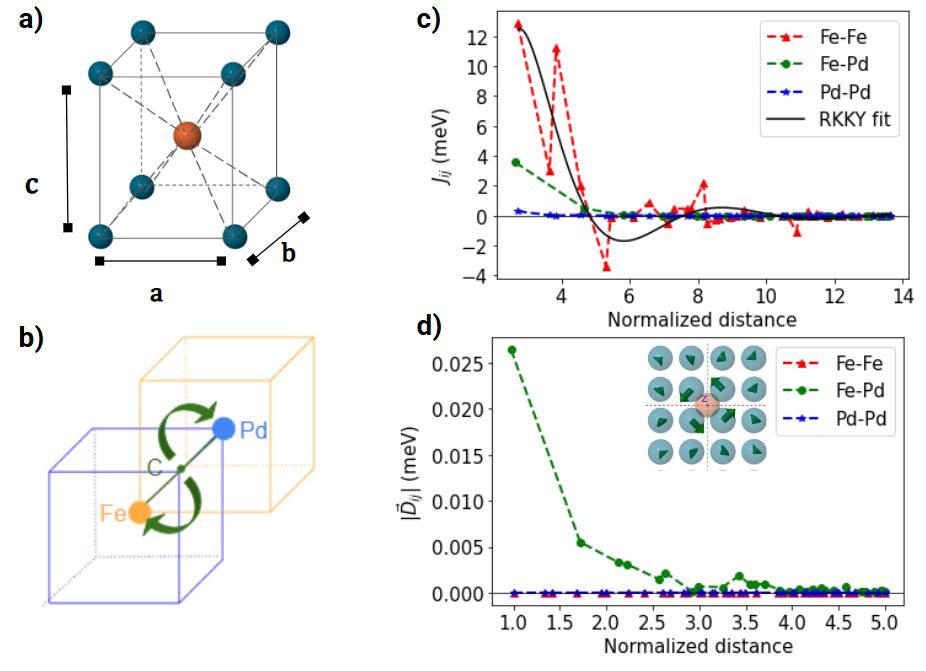}
    \caption{\textbf{Structural and theoretical magnetic interactions of pristine L$1_0$ FePd.} (a) Unit cell of FePd crystal (blue spheres: Pd; orange sphere: Fe), (b) application of the Moriya's rule in the Fe-Pd bond \cite{Moriya} (c) Exchange and (d) Dzyaloshinskii-Moriya coupling parameters for Fe-Fe (red dashed-lines), Fe-Pd (green dashed lines) and Pd-Pd (blue-dashed lines) interactions as a function of the normalized distance between the atomic pairs (in units of the lattice parameter). In (c), the solid black curve shows the RKKY fit.}
    \label{fig:FePd}
\end{figure}

Motivated by the experimental results, we shall theoretically explore some possible origins of chirality in FePd spin textures, starting from the electronic structure of the pristine material. Here, we model the L$1_0$ crystal structure with $a=b=2.722$ \AA\ and $c/a = 1.342$ \cite{FePd1, FePd2, FePd3, FePd4, FePd5, FePd6, FePd7}, as depicted in Figure \ref{fig:FePd}(a). In that configuration, both calculated magnetic moments and Curie temperature with various approaches (namely mean-field approximation, MFA, random-phase approximation, RPA, and Monte Carlo, MC) are shown in Table \ref{table1}. Other electronic structure results, such as the local density of states (LDOS), are presented in the Supplementary Note 1. As can be seen from Table  \ref{table1}, both computed properties are in good agreement with available experimental and theoretical results from the literature, as well as with the experimentally measured values for samples i.Nb/FePd and ii.FePd from Section \ref{sec:Experimental-Results} via magnetization measurements. 
%\IMans{Ivan: Annika, could you just mention here what technique you used to measure the magnetic moments in the samples? Ex.: XMCD.}\textcolor{red}{Its mentioned in section I before Table I}.

\begin{table*}[htb!]
    \centering
    \caption{\textbf{Magnetic properties of ordered L$1_0$ FePd and of the experimental samples investigated in this work, together with representative literature values.} The experimental magnetic moments reported here were obtained from the total magnetization of each full sample stack and are given as average moments per formula unit. Therefore, they include the combined contribution of all layers and atomic species.}
   % \textcolor{blue}{The experimental moments here refer to Fe+Pd or Fe moments? spin+orbital moments or spin? } \textcolor{red}{Annika: they are measured as average of the total magnetization from the whole sample system, including all layers and the response from all atoms together. Fe contributes most, but more generally Fe+Pd.}
    \begin{tabular}{@{}lccc@{}}
        \toprule
        \textbf{Source} & \textbf{Spin moment ($\mu_B$)} & \textbf{Orbital moment ($\mu_B$)} & \textbf{T$_C$ (K)} \\ 
        \midrule
        \textbf{This work: ord. L$1_0$ (theory)} & Fe: 2.83, Pd: 0.40 & Fe: 0.07, Pd: 0.03 & 629 (RPA)/888 (MFA)/630 (MC)\\
        \textbf{This work: i.Nb/FePd (expt.)} & \multicolumn{2}{c}{$3.2(0.2)$ $\mu_B$/f.u.} & -- \\
        \textbf{This work: ii.FePd (expt.)} & \multicolumn{2}{c}{$3.11(0.03)$ $\mu_B$/f.u.} & -- \\
        Ref. \cite{FePd1}: ord. L$1_0$ (theory) & Fe: 2.94, Pd: 0.29 & Fe: 0.06, Pd: 0.03 & -- \\ 
        Ref. \cite{FePd1}: ord. L$1_0$ (expt.) & -- & -- & 757 \\ 
        \bottomrule
    \end{tabular}
    \label{table1}
\end{table*}

Figure \ref{fig:FePd}(b-d) shows the application of Moriya's rules in the Fe-Pd bond \cite{Moriya}, calculated exchange and DMI couplings between Fe-Fe, Fe-Pd, and Pd-Pd as a function of the pairwise distance, as well as a schematic representation of the Fe-Pd DMI \textit{ab-initio} directions. Regarding the exchange couplings, we can observe that Fe-Fe interactions are stronger and exhibit magnetic frustration, especially at the $5^{th}$ neighboring shell. A characteristic oscillatory behavior of the Ruderman-Kittel-Kasuya-Yosida (RKKY) \cite{Ruderman1954} nature of exchange interactions in metals is obtained in Fig. \ref{fig:FePd}(c), similar to what is reported to bcc Fe \cite{Pajda2001,Kvashnin2016} and usual to noble-metal-$3d$-metal systems \cite{Mydosh1974}. 
%The interactions in the third neighbors seems to be more intense than in the second neighbors, which could be explained because the structure is not completely cubic ($c > a,b$).
In turn, the inversion symmetry of ideal $L1_0$ crystals is expected to lead to vanishing DMI, as can be seen in the case of Fe-Fe and Pd-Pd couplings (see Fig. \ref{fig:FePd}(d)). However, \textit{a priori}, Fe-Pd interactions present non-zero atomistic DMI values, which can be understood through Moriya's rules \cite{Moriya}. This is schematically shown in Fig. \ref{fig:FePd}(b), where the point bisecting the straight line between Fe and Pd does not constitute an inversion center ($\left|\mathbf{D}_{ij}\right|\neq 0$).%the first Moriya's rule says that if C is an inversion center, then $\mathbf{D}_{ij} = \mathbf{0}$, which is clearly not the case.
 When transforming it to the micromagnetic DMI stiffness tensor, $\mathcal{D}$ (see Supplementary Note 2), the translational symmetry of the Fe-Pd pairs nulls all its components, indeed vanishing any DMI contribution to the magnetic energy.
% \textcolor{blue}{Comment Annika: In this sentence and throughout the next sections you quite often have marked some text parts in red. These are places where you still insert the missing reference and/or information?} \IMans{Ivan: Yes, since supplementary sections can be modified during the finalization of the article, I only marked the parts that should be included when the text is ready. Otherwise, we would lose track of where to make the changes at the end.}

\subsection{Intermixing states: proof-of-concept in a more localized perspective}
\label{sec:a-localized-perspective}

%Even though the structures were deposited by MBE, atomic intermixing can happen during ...

%Studies show that the presence of \textit{intermixings} states within a multilayer system results in the contribution of DM interactions and the formation of non-collinear spin structures \cite{Pamela}. 
As in the pristine case we obtain the expected absence of contribution from the DMI, a first hypothesis for the existence of a global chirality, as demonstrated by the experiments in Section \ref{sec:Experimental-Results}, is the presence of atomic intermixing (as $S\neq 1$, see Table \ref{tab:Ku-S-parameters} and Fig. \ref{fig:XRD-HR-TEM}). This hypothesis is also motivated by the fact that, recently, the Co/Pd intermixing was identified as a cause of the emergence of isolated skyrmions in sputtered heterostructures \cite{Pamela}, by a local enhancement of the DM interaction strength.

Thus, we first simulate, via electronic structure calculations, localized intermixing states that are propagated throughout the structure periodically. 
%Thus, we can construct intermixing states from larger unit cells, called \textit{supercells}, which allow us to analyze the influence of these impurities in their surroundings. 
The intermixing states were separated into 9 main cases, as shown in Supplementary Note 2, where 4 are also shown in Figure \ref{fig:mainresults} (a), starting with two cases that seek to propagate the impurities in the $\hat{\mathbf{z}}$ direction (denoted by cases (\textit{i}) and (\textit{ii})) and another two that seek to propagate the impurities in the $\hat{\mathbf{x}}$ direction (cases (\textit{iii}) and (\textit{iv})), allowing us to explore different forms of local structure symmetry breaking.

\begin{figure*}[t!]
    \centering
    \includegraphics[width=0.7\linewidth]{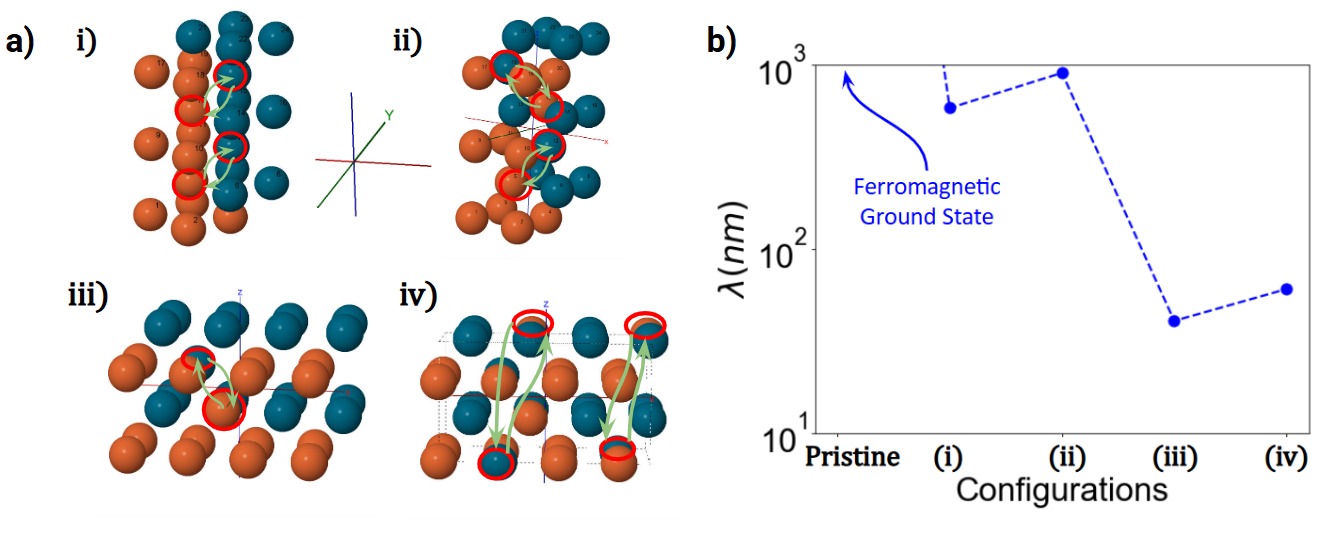}
    \caption{\textbf{Localized intermixing cases and screening indicators of disorder-induced noncollinearity.} (a) Types of atomic intermixings considered for calculations. Supercells (\textit{i}) and (\textit{ii}) grow in the $\hat{\mathbf{z}}$ direction, while in cases (\textit{iii}) and (\textit{iv}) they grow in the $\hat{\mathbf{x}}$ direction. (b) Characteristic modulation length $\lambda$ calculated from the stiffness and spiralization tensors ($\mathcal{A}$ and $\mathcal{D}$), as defined in the Supplementary Note 2. 
    %(d) Visualization of the spin spirals structure for case (\textit{iii}) via Jmol.
    %\cite{jmol} \textcolor{blue}{(J$_{\text{mol}}$?)}. 
    %e) Distribution of the $\hat{x}$, $\hat{y}$ and $\hat{z}$ components for the normalized magnetic moments for an internal layer. 
    In (a), green arrows indicate atom exchanges from the pristine structure.}
    \label{fig:mainresults}
\end{figure*}

\begin{figure*}[t!]
    \centering
    \includegraphics[width=\linewidth]{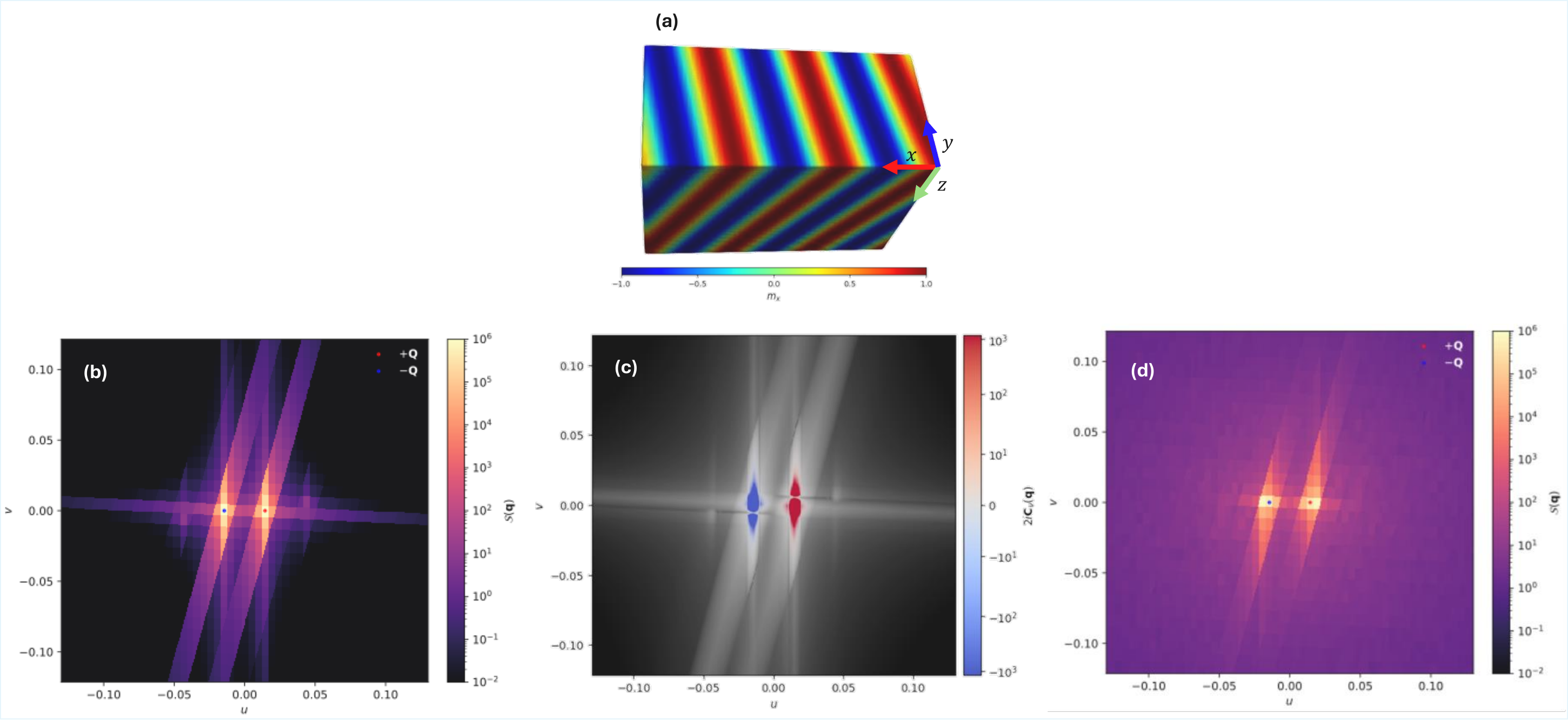}
    \caption{\textbf{Analysis of the spin texture for case (\textit{iii}).} 
    (a) Three-dimensional view of the real-space magnetization corresponding to a numerical energy minimum, showing the $\hat{\mathbf{x}}$ component. The supercell dimensions are $67.51\times33.75\times45.29$ nm$^3$. 
    (b) Static spin-spin structure factor $\mathcal{S}(\mathbf{q})$ at zero temperature for the configuration in (a). The figure shows a cut in reciprocal space containing both the ordering wave vectors $\mathbf{q}=\pm\mathbf{Q}$ (dominant Bragg peaks) and the $\Gamma$ point, with $\mathbf{q}=u\hat{\mathbf{u}}+v\hat{\mathbf{v}}$, where $\hat{\mathbf{u}}=\hat{\mathbf{Q}}$ and $\hat{\mathbf{v}}=(\hat{\mathbf{u}}\times\hat{\mathbf{x}})/|\hat{\mathbf{u}}\times\hat{\mathbf{x}}|$. $u$ and $v$ are given in reciprocal lattice units (r.l.u). (c) Chirality measure (spin asymmetry along polarization axis $\boldsymbol{\nu}$), $2iC_{\boldsymbol{\nu}}$ (Eq.~\ref{eq:asymmetry}), for $\boldsymbol{\nu}=\mathbf{P}_z=[001]$, with the structure-factor intensity shown in grayscale in the background. (d) Thermal average of $\mathcal{S}(\mathbf{q})$ at $T=300$ K, obtained from stochastic spin-dynamics simulations over $N=150$ snapshots taken at every $\Delta t=100$~fs.}
    \label{fig:analysis-magnetization}
\end{figure*}

To evaluate whether localized intermixing can promote chiral noncollinearity, we analyzed the spiralization tensor $\mathcal{D}$ and the characteristic modulation length $\lambda$, as defined in Supplementary Note 2. At this stage, $\lambda$ should be viewed only as an indicator of the leading finite-$\mathbf{q}$ instability within the reduced flat spin-spiral ansatz, and not as a direct prediction of the full real-space magnetic pattern. The results are summarized in Fig. \ref{fig:mainresults}(b). As expected, the pristine L$1_0$ structure gives $\mathcal{D}=0$, and therefore $\lambda\rightarrow\infty$, consistent with a ferromagnetic ground state in this simplified picture. Cases (\textit{i}) and (\textit{ii}) yield only weak effective spiralization, with very long characteristic wavelengths, $\lambda\sim582$ nm and $\lambda\sim903$ nm, respectively. This points to only a very weak finite-$\mathbf{q}$ tendency, i.e., an almost collinear state on the scale of the present supercells. By contrast, cases (\textit{iii}) and (\textit{iv}) show larger components of $\mathcal{D}$, and the corresponding wavelengths are reduced to $\lambda\sim40.6$ nm and $\lambda\sim60.4$ nm, respectively. These values place the modulation in the nanometer range and identify these cases as the most promising ones for explicit atomistic calculations.

%\IM{[Juan: depois vamos trocar essas figuras.]}
Accordingly, Figure \ref{fig:analysis-magnetization}(a) shows the obtained ground state for case (\textit{iii}), where we observe the formation of a noncollinear spin structure. From the calculated real-space configurations, however, it is difficult to determine by visual inspection whether a net chirality is present. To address this, chirality can be assessed both quantitatively and qualitatively using two complementary measures: (\textit{1.}) the effective presence of a Néel-Bloch-type wall admixture, quantified by the scalar $\bar{C}_N$ defined in the Supplementary Note 3, in which a value $0<\bar{C}_N<1$ reflects the presence of a Néel-Bloch admixture (see discussion in Supplementary Note 3), and a nonzero Bloch-type wall component signals magnetic chirality in a geometrical sense \cite{Cheong2022}, as such wall types break all mirror symmetries; and (\textit{2.}) the vector quantity $C_{\boldsymbol{\nu}}$ defined in Eq.~\ref{eq:asymmetry}, which enables a direct qualitative comparison with experimental GISANS results. Before proceeding, it is important to note that although the measures in (\textit{1.}) and (\textit{2.}) are not identical, they are closely related. The GISANS spin-flip asymmetry (Eq.~\ref{eq:asymmetry}) probes the projected, momentum-space net chirality of the texture, whereas the geometrical definition in the Supplementary Note 3 -- that motivates to compute $\bar{C}_N$ -- captures chirality in real space. Their connection becomes transparent in the limit of a single-$\mathbf{q}$ Bloch state with net handedness, where both quantities become equivalent \footnote{This can be seen by considering a flat-spiral \textit{ansatz} $\mathbf{m}(\mathbf{r}) = m_0\!\left[\hat{\mathbf{e}}_1 \cos(\mathbf{q}_0\!\cdot\!\mathbf{r}) + \hat{\mathbf{e}}_2 \sin(\mathbf{q}_0\!\cdot\!\mathbf{r})\right]$, with $\hat{\mathbf{e}}_1 \perp \hat{\mathbf{e}}_2$, $\hat{\mathbf{e}}_{1,2}\perp \hat{\mathbf{q}}_0$. Evaluating $\mathbf{C}$ (Eq.~\ref{eq:asymmetry}) and the $\alpha$-component of the chirality tensor $\boldsymbol{\mathcal{L}}_\alpha = (\partial_\alpha \mathbf{m}) \times \mathbf{m}$ (see Supplementary Note 3), one finds, up to a proportionality constant, $\mathbf{C}(\mathbf{q}_0)\propto \mathbf{K}_\alpha(\mathbf{q}=0)$, where $\mathbf{K}_\alpha$ is the Fourier transform of $\boldsymbol{\mathcal{L}}_\alpha$. Explicitly, $\mathbf{K}_\alpha(\mathbf{q}=0) = -2q_{\alpha,0}i\,\mathbf{C}(\mathbf{q}_0)$, showing that the zero-momentum value of $\mathbf{K}_\alpha$ measures twice the spiral wavevector times the chiral vector $i\mathbf{C}$ at $\mathbf{q}_0$, with the factor 2 reflecting the equal-weight $\pm \mathbf{q}_0$ modes.}.
%as shown in . 

%%%%% COMENTEI POIS MUDAMOS A ANÁLISE
%\textcolor{red}{Here we are dealing with case (iii), where, from the x, y, and z components of the magnetic moments, as seen in \ref{fig:mainresults}, it is possible to observe that the rotation of the spins occurs mostly in the xy plane, clearly presented in figure \ref{fig:mainresults} (d). Not only that, but with the propagation direction occurring in the x direction, we have the spiral behaving more similarly to a N\'eel-type spiral.}

From the spin structure shown in Fig. \ref{fig:analysis-magnetization}(a), we estimate the zero-temperature, zero-field values of both $\bar{C}_N$
%\textcolor{blue}{(Annika: both C$_N$ of what? Why two?)} \textcolor{green}{I think the 'both' is referring to $\bar{C}_N$ and the $\mathcal{S}(\mathbf{q})$} \IMans{Ivan: Yes, it is just the order of the phrase.}
and the Bragg peak of the static spin-spin structure factor $\mathcal{S}(\mathbf{q})$, which defines the ordering wave vector of the FePd intermixed model system. For this spin structure, 
%\textcolor{blue}{(Annika: Which case? For the spin spiral state related to case III?)} \IMans{Ivan: changed to 'spin structure' to be more understandable.}
we obtain $\bar{C}_N \sim 0.78$ and $\mathbf{Q} = (0.277,\,0.016,\,0.138)$ nm$^{-1}$ for a large supercell containing $62^3$ units of 32 atoms ($\sim 7.6\times10^6$ atoms in total). This corresponds to a characteristic in-plane modulation period of $\lambda = 2\pi/|\mathbf{Q}_{\parallel}| \sim 22.6$ nm, which is smaller than the experimental range of $2\pi/Q_y\sim 185$-$217$ nm (Table \ref{tab:Q-angles}). The discrepancy is expected, given the simplicity and periodicity of the present intermixing model. As also expected, the noncollinear classical ground state is not described by a single spiral with ordering wave vectors $\mathbf{q} = \pm\mathbf{Q}$, but can be expressed as a superposition of $k$ spirals with distinct wave vectors $\pm\mathbf{Q}_k$. However, the intensity of $\mathcal{S}(\mathbf{Q}_k)$ is at least $\sim 10^3$ times smaller for $\mathbf{Q}_k \neq \mathbf{Q}$ and rapidly decays, making $\mathbf{Q}$ the dominant Bragg peak. The resulting spin texture, together with the corresponding $\mathcal{S}(\mathbf{q})$, is shown in Figs. \ref{fig:analysis-magnetization}(a,b). For remaining cases presented in Fig.~\ref{fig:mainresults} and Supplementary Note 2, the computed $\bar{C}_N$ values fall within the $0.48$-$0.82$ range.
%\textcolor{blue}{(Not readily written yet? Reference to above "different C$_N$-comment)}
%&, \IM{...} 
%As discussed in the \IM{Supplementary Note []}, a value of $\bar{C}_N < 1$ reflects the presence of a Néel-Bloch-type wall admixture 
%\textcolor{red}{Note Annika: as mentioned before, this should be coming earlier}.
This consistent deviation from 1 ($\bar{C}_N<1$) indicates that DMI, induced by chemical disorder, robustly drives the emergence of magnetic chirality across all tested intermixing patterns.

As mentioned before, a full characterization of chirality also requires evaluating the asymmetry $2iC_{\boldsymbol{\nu}}$ (Eq.~\ref{eq:asymmetry}) projected onto the polarization axis $\boldsymbol{\nu}$, which enables direct comparison with the GISANS measurements (see Fig. \ref{fig:analysis-magnetization}(c)). For Fig. \ref{fig:analysis-magnetization}(c), we choose the polarization vector $\boldsymbol{\nu}=\mathbf{P}_z=[001]$. From the obtained results, we observe a pronounced asymmetry around the ordering vectors $\pm\mathbf{Q}$, consistent with the symmetry relation $C_{\boldsymbol{\nu}}(\mathbf{Q}) = -C_{\boldsymbol{\nu}}(-\mathbf{Q})$, thereby demonstrating the emergence of chirality in reciprocal space and yielding an asymmetry that is qualitatively consistent with the GISANS measurements.
%\textcolor{red}{Note Annika: it is not clear here to me, which $Q_{\nu}$ you mean, the exp. observed one? Can you describe how you get to this result with one more sentence?}

Since the experimental results discussed in the previous sections are obtained at room temperature, it is important to assess the behavior of the magnetic configuration found under comparable conditions. To evaluate the long-time thermal stability of the noncollinear structures, we performed stochastic spin-dynamics at $T=300$\,K, saving snapshots every $\Delta t=100$\,fs and averaging the static structure factor over $N=150$ nearly independent configurations. As shown in Fig.~\ref{fig:analysis-magnetization}(d), the dominant Bragg-like peaks at $\pm\mathbf Q$ persist at 300\,K with unchanged positions within the reciprocal-space resolution, but broaden appreciably: the effective width increases from $\kappa(T=0)\sim0.01$-$0.017$ to $\kappa(T=300)\sim0.016$-$0.021$, implying a $\sim15$-$45\%$ reduction of the correlation length. In contrast, weaker satellite peaks are not resolved above the local background within our analysis window, consistent with thermally induced dephasing that redistributes the spectral weight into a diffuse component around $\pm\mathbf{Q}$. The robustness of the principal peaks is consistent with strong Fe-Fe exchange and a Curie temperature exceeding room temperature.

\subsection{Interdiffusion with \textit{ab initio} accuracy: disorder- and gradient-induced net global chirality}
\label{sec:gradient-interdiffusion}

%\textcolor{red}{General Note Annika: Can any of the further text go into the methods section or supplementary? I feel its a very long description with only a very short results part at the end. Maybe its ok if it stays as-is for the first submission, but might give better overview + reading.}
Even though the model of Sec.~\ref{sec:a-localized-perspective} provides a clear proof-of-concept for the emergence of chirality through atomic intermixing in the nominally centrosymmetric L$1_0$ FePd -- whose results incorporated both short- (i.e., within the nearest neighborhood) and long-range induced or modified magnetic interactions ($J_{ij}$ and $\mathbf{D}_{ij}$) 
%\textcolor{blue}{(Annika: Where are short- vs. long-range interactions described? It is not clear to me how you differ here between the "short" and "long-range" interactions and where in the discussion above is mentioned about which?)} \IMans{This is a good question. Here, short and long-range interaction mean that we are not considering only nearest neighbors or changes in the nearest neighborhood, but an extended range of interactions. Long-range may be important also for the most energetically favorable magnetic configuration. I will change the text accordingly.}, 
-- it also presents at least two limitations: (\textit{i}) the supercell employed is relatively small compared to the real sample and repeats periodically in space, thereby introducing an artificial periodicity of the disorder; and (\textit{ii}) chirality and the formation of noncollinear ground states depend sensitively on the details of the intermixing model. These considerations advise against relying solely on direct DFT calculations to model the experimental system in this case, even with approaches such as the well-known special quasirandom structures \cite{Zunger1990}, since reproducing the radial correlation functions of random alloys does not ensure a correct description of long-range magnetic interactions, nor does it capture structural features of real samples, such as gradient interdiffusion.  

In contrast, the GISANS measurements presented in Sec.~\ref{sec:GISANS} reveal a chirality that emerges on a \textit{global} scale, together with clear evidence of gradient interdiffusion throughout the sample (see Fig.~\ref{fig:XRD-HR-TEM}). To address these challenges, we adopt a model that bridges the atomistic and mesoscopic scales, inspired by Ref.~\cite{Gradient}, which explicitly incorporates gradient interdiffusion: the gradient model (see Supplementary Note 4). In this framework, the distribution of atomic intermixing across the material generates a compositional gradient along the $\hat{\mathbf{z}}$ direction. In the present work, however, we introduce two key modifications to the original formulation: (\textit{i}) we consider a gradient involving two distinct sublattices; and (\textit{ii}) we establish a direct link to realistic \textit{ab initio} results, rather than relying on the Fert-Levy model~\cite{Fert1980} with phenomenological or semi-empirical parametrizations. In particular, the explicit two-sublattice formulation (modification (\textit{i})) allows independent gradients $k^{(1)}\neq k^{(2)}$ and central stoichiometries $\bar{x}^{(1)}\neq \bar{x}^{(2)}$, which is relevant to represent ordered L1$_0$ alloys under intermixing. In turn, regarding modification (\textit{ii}), the connection is achieved through a multitask graph neural network, \texttt{SAGNN}. Details of the neural network, including its architecture and hyperparameters, are provided in Sec.~\ref{sec:methods} and in the Supplementary Note 5. As such, the model is designed to capture both the qualitative and quantitative aspects of chirality induced by the DMI arising from gradient intermixing on a global scale. The surrogate was selected based on validation performance and its ability to reproduce the rare large-amplitude interactions most relevant to finite-$\mathbf{q}$ instabilities; extended benchmarking is provided in the Supplementary Note 5.

\begin{figure*}[t!]
    \centering
    \includegraphics[width=\linewidth]{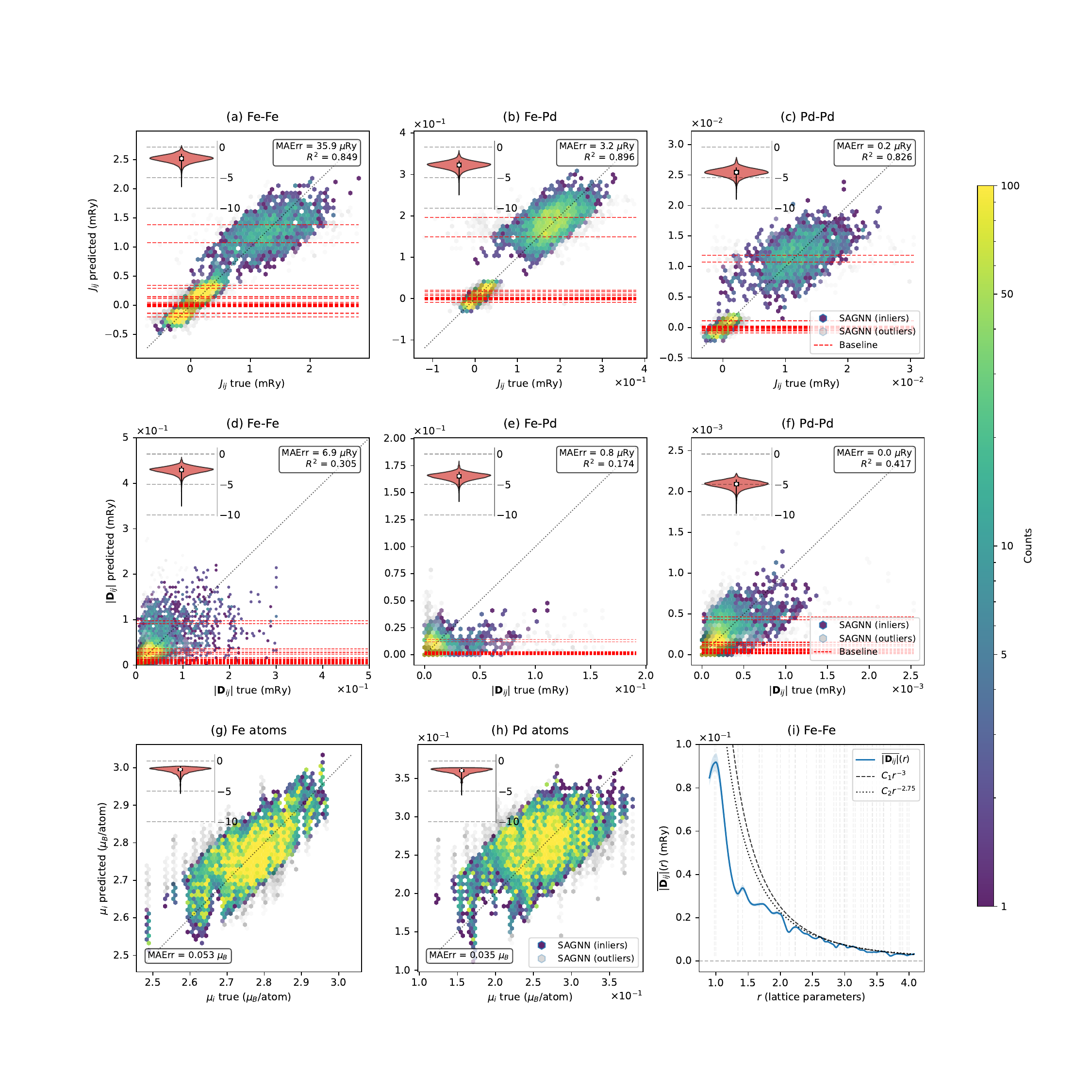}
    \caption{\textbf{Parity and radial decay plots for magnetic interactions and spin moments.} 
    (a-c) Fe-Fe, Fe-Pd, and Pd-Pd exchange interactions; (d-f) DMI; (g,h) spin magnetic moments; (i) radial profile of the Fe-Fe DMI magnitude $\overline{|\mathbf{D}_{ij}|}(r)$. Each plot compares the fully \textit{ab-initio} (true) values, obtained from DFT, with the corresponding \texttt{SAGNN} predictions across the validation set. 
    Gray hexagons indicate points lying outside the 90\% prediction interval, while the dotted red lines in panels (a-f) denote the baseline values $\bar{J}_{ij}^{\text{base}}(s,p) = \frac{1}{n(\mathcal{E}_{s,p}^{\text{train}})} \sum_{(i,j)\in\mathcal{E}_{s,p}^{\text{train}}} J_{ij}$ 
    and $|\bar{\mathbf{D}}_{ij}|^{\text{base}}(s,p) = \frac{1}{n(\mathcal{E}_{s,p}^{\text{train}})} \sum_{(i,j)\in\mathcal{E}_{s,p}^{\text{train}}} |\mathbf{D}_{ij}|$ 
    (see Methods), computed as shell- $s$ and pair-type $p$ averages in the training set. 
    The violin plots in the insets show the absolute error distributions (in mRy) on a logarithmic scale; the internal box denotes the first and third quartiles, and the central white dot marks the median. 
    The dotted black lines represent perfect predictions. 
    The coefficients of determination ($R^2$) and mean absolute errors (MAErr) are reported for each plot.}
    \label{fig:parity-plots-main}
\end{figure*}

\begin{figure*}[t!]
    \centering
    \includegraphics[width=\linewidth]{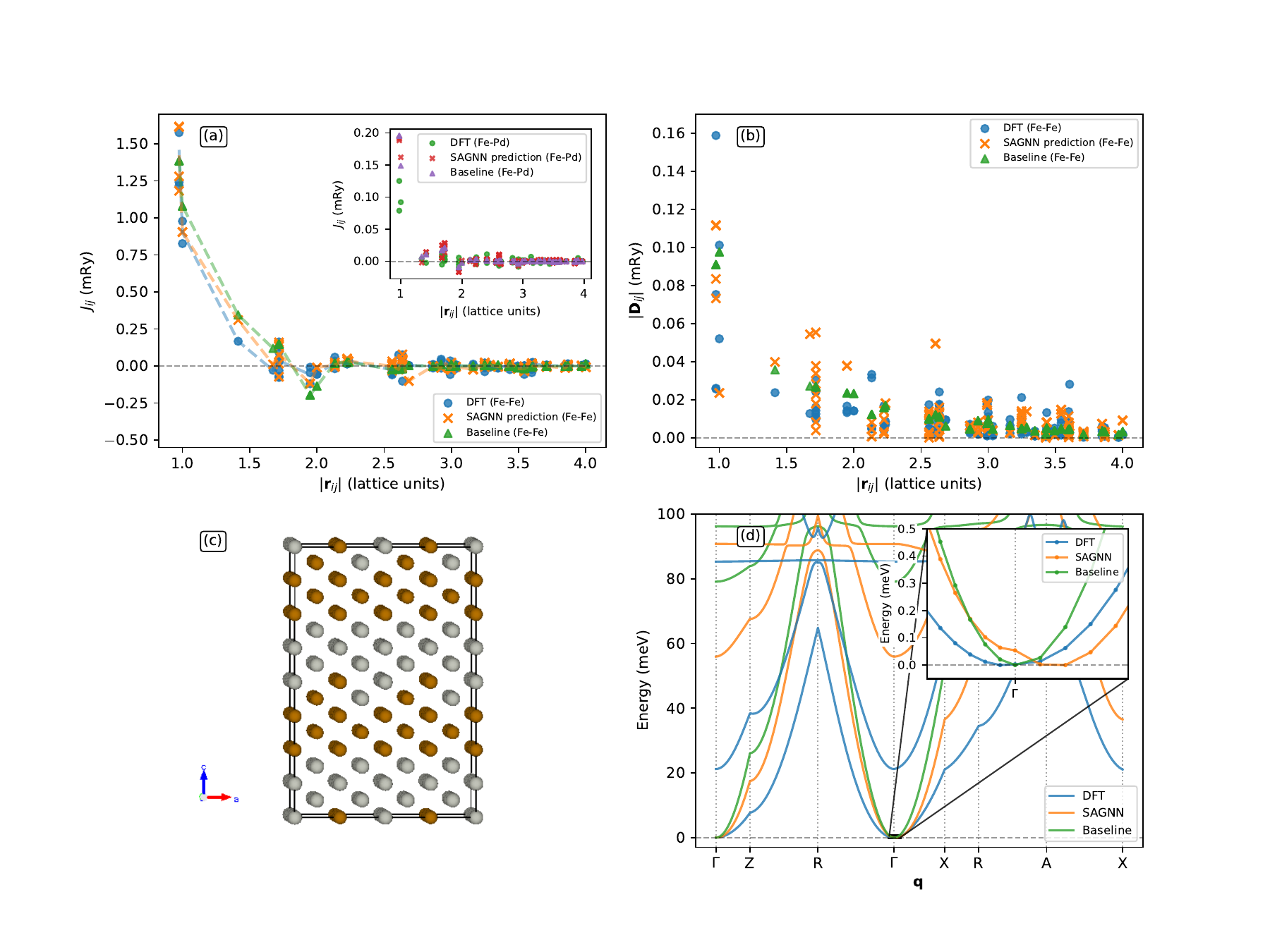}
    \caption{\textbf{Magnetic interactions in a disordered Fe$_{50}$Pd$_{50}$ supercell mimicking a Fe-Pd gradient and their associated properties.} 
    (a) Fe-Fe exchange interactions and Fe-Pd interactions (\textit{inset}); 
    (b) Fe-Fe DMI interactions; 
    (c) atomic structure of the disordered FePd, modeled as a random $4\times 2\times 2$ supercell (32 atoms), constructed to mimic a Fe-Pd compositional gradient and lying outside the \texttt{SAGNN} training and validation sets; 
    (d) softest branches of the adiabatic magnon spectrum obtained using the three interaction sets (DFT, \texttt{SAGNN}, and baseline). All interactions are calculated with respect to a single Fe reference site, randomly selected among the 16 Fe atoms in the cell. 
    Blue dots denote direct DFT results, while orange crosses (green triangles) represent the corresponding \texttt{SAGNN} (baseline) predictions. 
    In the \textit{inset} of panel~(a), green dots indicate DFT Fe-Pd interactions, and red crosses (purple triangles) the corresponding \texttt{SAGNN} (baseline) predictions. 
    The dotted lines in panel~(a) serve as guides to the eye and indicate the average exchange interaction within each coordination shell for each model. 
    In panel~(c), orange (gray) spheres represent Fe (Pd) atoms. 
    The \textit{inset} in panel~(d) shows a zoom of the magnon spectra around the $\Gamma$ point.}
    \label{fig:jij-net-comparison}
\end{figure*}

\subsubsection{Modeling via deep learning}

Figure \ref{fig:parity-plots-main}(a-h) shows the parity plots for the obtained $J_{ij}$, $|\mathbf{D}_{ij}|$, and $\mu_i$ and the corresponding error distributions (\textit{insets}). The first aspect to note is that the network captures a much more complex scenario than the baseline model (see Methods), which produces, by construction, a series of discrete values (plateaus) for each pair type on each neighboring shell. For instance, the network captures a much wider range of $J_{ij}$ and $|\mathbf{D}_{ij}|$ interaction values -- with compatible amplitude to the validation dataset -- and deeper antiferromagnetic Fe-Fe exchange interactions, thus learning a RKKY-like behavior. This can be further attested in the example Fig. \ref{fig:jij-net-comparison}, which, also to assess generalization beyond the validation dataset, compares the DFT-computed $J_{ij}$ (Fe-Fe, Fe-Pd) and $|\mathbf{D}_{ij}|$ (Fe-Fe) couplings in a disordered Fe$_{50}$Pd$_{50}$ alloy -- modeled as a random, out-of-sample
%\textcolor{blue}{(Annika: what does that mean? )} \IMans{Ivan: out-of-sample means that the alloy model used is not in the training/validation sets, in other words, the network has never seen it.} 
$4\times2\times2$ supercell (32 atoms) -- with the corresponding \texttt{SAGNN} predictions. As can be seen in Fig. \ref{fig:jij-net-comparison}(c), the alloy supercell model is constructed to mimic a relevant case for the present investigation: a Fe-Pd gradient. The plots in (a) and (b) display all interactions involving a given Fe reference site, displayed as a function of the pairwise distance $|\mathbf{r}_{ij}|$. The results show that the neural network not only reproduces the relevant negative (antiferromagnetic) couplings, but also captures the variability of interaction strengths within each coordination shell -- an aspect that the baseline model, by construction, cannot represent. In addition, it reproduces the long-range oscillatory behavior characteristic of RKKY-type interactions, preserves the correct order of magnitude across shells, and can resolve stronger nearest-neighbor couplings (in closer agreement with the DFT data), whereas the baseline remains constrained to shell-averaged values.

Beyond a pure interaction-level analysis, also interesting is to characterize how these predicted magnetic couplings perform under selected experimentally relevant observables. In order to do that, we used the interactions directly predicted from $\texttt{SAGNN}$ in spin dynamics calculations at zero or finite temperatures (see Methods), comparing with the results coming from the baseline model and direct DFT calculations, under the same conditions. One observable that can be directly accessed is the transition temperature. For the supercell under analysis, the DFT+\texttt{UppASD} calculation identified a noncollinear ground-state spin structure, and the transition temperature was obtained via Monte Carlo and specific heat analysis \footnote{The specific heat at constant volume was calculated as the change in specific internal energy ($e$) with respect to temperature at constant volume $c_v=\left(\frac{\partial e}{dT}\right)_v$.} as 576 K. For the \texttt{SAGNN} and baseline models, using the same method, we obtained 786 K and 996 K, respectively. Another observables are the ordering channel and aspects of the magnon spectrum, which are shown in Fig. \ref{fig:jij-net-comparison}(d). From those results, we see that the baseline model predicts the wrong instability (FM instead of finite-$\textbf{q}$), and the neural network partially corrects that by predicting also a noncollinear ground-state, representing a qualitative gain over the baseline. Moreover, we see from the adiabatic magnon spectra, in the softest branches, that the neural network predicts a wider shape in the neighborhood of the $\Gamma$-point, associated with a smaller long-wavelength stiffness from the baseline, towards the obtained directly with DFT. This is well-correlated with the smaller transition temperature found for the \texttt{SAGNN} model in comparison with the baseline. This comparison is important not only at the level of prediction error, but also at the level of the emergent magnetic physics. The shell-averaged baseline suppresses the local environment fluctuations induced by chemical disorder, precisely relevant ingredients to frustration, finite-$\mathbf{q}$ instabilities, and the emergence of chirality. The fact that the baseline predicts the wrong ordering tendency, whereas \texttt{SAGNN} recovers a noncollinear finite-$\mathbf{q}$ state in this example, indicates that preserving local interaction variability is not a secondary refinement, but a necessary condition for realistic microscopic description of disorder-induced chirality in FePd.

The average execution time per exchange interaction was $0.92\pm0.07$\,ms on an 8-core MacBook Air M3, representing a speed-up of three to four orders of magnitude compared with direct DFT calculations using the method of Ref.~\cite{Frota-pessoa2000}, even when accounting only for the required post-processing steps (i.e., assuming a well-converged electronic structure).
 %\IM{[Comentar aqui mais sobre a figura, e a diferença de tempo para se obter os dados com ML.]} 
Since the graphs in \texttt{SAGNN} are divided by shell and pair types, it successfully captures the magnetic interactions in all three orders of magnitude: Fe-Fe $>$ Fe-Pd (mixed) $>$ Pd-Pd.
%, as detailed by \IM{Supplementary Figure []}. 
The obtained mean absolute errors for exchange couplings are 35.9 $\mu$Ry (Fe-Fe), 3.2 $\mu$Ry (Fe-Pd), and 0.2 $\mu$Ry (Pd-Pd), which are comparable to errors arising from the approximation made within the DFT method used (see Section \ref{sec:methods}). These observations extend to the DMI predictions. Although the Pd-Pd $\mathbf{D}_{ij}$ interactions are  weak -- on the order of a few $\mu$Ry --, the network successfully captures them due to the shell- and pair-specific standardization procedure applied during training (see Methods).

%however with some caveats: Pd-Pd $|\mathbf{D}_{ij}|$ interaction magnitudes are very small, on the order of a few $\mu$Ry, 

%%% Já descrito antes
%This approach enables a full translation of first-principles atomistic data to a larger, mesoscopic scale -- where direct DFT simulations become computationally prohibitive.
%Details of the model are given in the Supplementary Material and in the original article \cite{Gradient}, but it is important to highlight here the case of our FePd system.
%The model arises as a qualitative analysis to explain how DMI surges and behaves throughout intermixed systems. 

\subsubsection{Construction and analysis of the mesoscopic FePd model}

To construct the mesoscopic model of the FePd thickness in the FePd/Nb system, we employed the modified version of the gradient model (MGM), as described in detail in Supplementary Note~4, aiming to reproduce, as closely as possible, the observed characteristics of samples i.Nb/FePd and ii.FePd. Since some properties of the FePd thickness were experimentally determined -- namely the long-range order parameter $S$ (see Table~\ref{tab:Ku-S-parameters}) and an overall Fe concentration of approximately $50\%$ ($c_{\textnormal{Fe}} \sim 0.5$) -- we use the pair $(S, c_{\textnormal{Fe}})$ to define the mean occupation probabilities $p^{(l)}_{\text{Fe}}$ (with $p^{(l)}_{\text{Pd}}=1-p^{(l)}_{\text{Fe}}$) and the central plane stoichiometries  $\bar{x}^{(l)}$ of each sublattice $l=1,2$. In addition, the slopes $k^{(l)}$ control both the magnitude and the sign of the compositional gradient in each sublattice; by construction, although some freedom exists in choosing $k^{(l)}$, the allowed range is tightly constrained by the values of $\bar{x}^{(l)}$ and, consequently, by the experimentally determined $(S, c_{\textnormal{Fe}})$. For a binary alloy, the long-range order parameter $S$ is here defined as~\cite{Bragg1934}

\begin{equation}
S = \frac{p^{(1)}_{\text{Fe}} - p^{(2)}_{\text{Fe}}}{2\min(c_{\text{Fe}}, 1 - c_{\text{Fe}})},
\end{equation}

\noindent to which we assign the values listed in Table~\ref{tab:Ku-S-parameters}. Furthermore, it is reasonable to impose a higher Pd concentration near the substrate interface, gradually converging toward Fe$_{50}$Pd$_{50}$ away from the Pd host, consistent with Fig.~\ref{fig:XRD-HR-TEM}(a). Following these considerations, we construct a supercell with the L1$_0$ crystal structure and $n_x =  n_y = 50a$ and $n_z =27c$ ($\sim13.6\times 13.6\times 9.9$ nm$^3$), containing 135,000 atoms; the resulting system is shown in Supplementary Figure 3 and the source code to produce the structure following the MGM is freely available in Zenodo. Together with a neighbor cutoff radius of 4 lattice parameters -- which results in 388 interactions of each type per site in the supercell (see Methods) -- this amounts to a total of $\sim1.04 \times 10^8$ interactions (including both exchange and DMI) to be inferred by the neural network. For target values $S^{\text{target}} = 0.52$, $c_{\text{Fe}}^{\text{target}} = 0.45$ (close to $0.5$), and $(k^{(1)}, k^{(2)}) = (5\times10^{-4}, -10^{-2})$, we obtain $(\bar{x}^{(1)}, \bar{x}^{(2)}) = (0.684, 0.784)$, yielding effective $(S^{\text{eff}}, c_{\text{Fe}}^{\text{eff}})$ values within $\sim0.1\%$ of the target. The termination adjacent to the Pd host has an approximate composition of Fe$_{0.38}$Pd$_{0.62}$, while the opposite termination reaches Fe$_{0.52}$Pd$_{0.48}$, in full agreement with the initial assumptions.

For the constructed mesoscopic FePd supercell, a full real space relaxation involving $\sim 10^8$ interactions and 135,000 inequivalent sites -- e.g., via Monte Carlo schemes such as simulated annealing -- is memory-wise computationally challenging. We thus adopt a reciprocal space effective-medium approach by projecting the atomistic Hamiltonian onto a flat spin spiral \textit{ansatz}, with the total energy

\begin{equation}
\label{eq:total-energy-spiral-ansatz}
\begin{split}
E(\mathbf{q},\hat{\mathbf{n}})=-\sum_{\mathbf{r}}\left[\bar{J}(\mathbf{r})\cos(\mathbf{q}\cdot\mathbf{r})+\hat{\mathbf{n}}\cdot\bar{\mathbf{D}}(\mathbf{r})\sin(\mathbf{q}\cdot\mathbf{r})\right]\\
+E_{\text{ani}}(\hat{\mathbf{n}}),
\end{split}
\end{equation}

\noindent where $\mathbf{r}$ is a lattice vector, $E_{\text{ani}}$ the magnetic anisotropy energy term, $\bar{J}(\mathbf{r})=(1/N_{\mathbf{r}})\sum_{(i,j):\mathbf{r}_{ij}=\mathbf{r}}J_{ij}$, $\bar{\mathbf{D}}(\mathbf{r})=(1/N_{\mathbf{r}})\sum_{(i,j):\mathbf{r}_{ij}=\mathbf{r}}\mathbf{D}_{ij}$, and $N_{\mathbf{r}}$ is the number of pairs with separation $\mathbf{r}_{ij}=\mathbf{r}$. In this way, the disordered alloy is approximated by a translationally invariant virtual crystal, mapping the complex random interaction network onto a periodic effective lattice. The ground state is then obtained by minimizing $E(\mathbf{q},\hat{\mathbf{n}})$ without imposing any constraint on the relation between $\hat{\mathbf{n}}$ and $\mathbf{q}$, beyond the normalization $|\hat{\mathbf{n}}|=1$ (see Methods).

Regarding the magnetic anisotropy, fully disordered FePd ($S=0$) adopts the fcc (A1, austenite) parent structure rather than the tetragonal L$1_0$ phase~\cite{Laughlin2005}. In this limit, the magnetocrystalline anisotropy is orders of magnitude smaller than in fully ordered L$1_0$ FePd ($S=1$)~\cite{Kamp1999,Cui2004,PhysRevB.70.224408}. In the intermediate, partially ordered regime relevant here, the MAE is strongly correlated with the $S$ parameter and is theoretically described by a power law $\ E_{\mathrm{MAE}}\propto S^{n}$ with $n\sim 1.6$-$2.4$~\cite{Kota2012}. In our atomistic simulations we employ a homogeneous on-site anisotropy ($K_i\equiv K$, see Methods) and do not include long-range dipolar interactions or the explicit energetics of closure domains. Therefore, the anisotropy entering the model should not be interpreted as $K_u$ of the experimental sample, but rather as an \textit{effective} perpendicular anisotropy parameter of the reduced Hamiltonian that coarse-grains both the spatial inhomogeneity of the local MAE and energetic contributions not explicitly captured by the model (including magnetostatic/shape effects). In this sense, $K$ acts as a control parameter for the stability of the modulated state.

\begin{figure}[htb!]
    \centering
    \includegraphics[width=\linewidth]{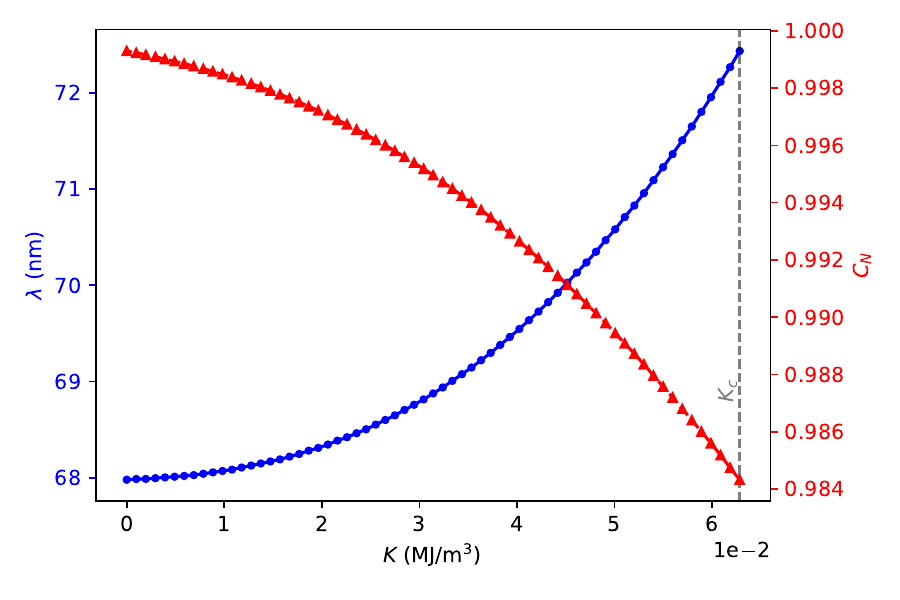}
    \caption{\textbf{Sensitivity analysis of the stability of the chiral magnetic ground state in the mesoscopic FePd model.}
    In-plane modulation period, $\lambda=2\pi/|\mathbf{Q}_{\parallel}|$, and chirality measure $C_N$ as functions of the effective anisotropy parameter $K$. The vertical dotted line marks the critical anisotropy value $K_c$.}
    \label{fig:sweep-K}
\end{figure}

Figure \ref{fig:sweep-K} shows the sensitivity analysis results with respect to $K$. A finite-$\mathbf{q}$ solution exists and is stable within an anisotropy window, collapsing into the homogeneous $\mathbf{Q}=0$ state above a critical value $K_c$. The characteristic in-plane modulation period $\lambda=2\pi/|\mathbf{Q}_{\parallel}|=2\pi/\sqrt{Q_x^2+Q_y^2}$ is found to be in the range $68$-$72.5$ nm. This represents a clear improvement over the localized model of Section~\ref{sec:a-localized-perspective} and brings the theoretical modulation period substantially closer to the experimentally estimated in-plane period of $185$-$217$~nm.  The chirality measure $C_N$ evolves continuously, showing a progressive deviation of the ground-state from the pure Néel limit. As a representative case within this stability window, for $K=0.06$ MJ/m$^3$ we find a minimum at $\mathbf{q}=\mathbf{Q}=(-0.088,\,0.007,\,-0.017)\,\text{nm}^{-1}$, with corresponding rotation axis $\hat{\mathbf{n}}_{\text{min}}$ %\textcolor{red}{Note Annika: Why is this Q so different to the exp. observed one? Later, you describe a comparison of lambda with the exp results which probably fits much better. Should this Q be mentioned here at all?}. 
From the definition $C_N=1-(\hat{\mathbf{n}}_{\text{min}}\cdot\hat{\mathbf{Q}})^2$, we obtain $C_N\sim0.9856$, indicating a Néel-Bloch admixture and, therefore, the emergence of magnetic chirality driven by disorder-induced DMI. For this value of $C_N$, the effective DMI tilts the rotation axis by only a few degrees ($\sim 6.9^\circ$) away from the pure Néel configuration, consistent with GISANS measurements that also indicate a relatively weak but finite asymmetry for sample ii.FePd. This conclusion is further supported by the asymmetry $E(\mathbf{Q},\hat{\mathbf{n}}_{\text{min}})\neq E(-\mathbf{Q},\hat{\mathbf{n}}_{\text{min}})$. The wavevector $\mathbf{Q}$ also gives a characteristic in-plane modulation period of $\lambda\sim71.2$\,nm. Regarding the asymmetry defined in Eq.~\ref{eq:asymmetry}, for a single-$\mathbf{q}$ \textit{ansatz} the quantity $C_{\boldsymbol{\nu}}$ is proportional to the factor $(\mathbf{Q}\cdot\hat{\mathbf{n}}_{\text{min}})(\mathbf{Q}\cdot\boldsymbol{\nu})$. Since $(\mathbf{Q}\cdot\hat{\mathbf{n}}_{\text{min}})\neq 0$, this configuration is also chiral according to the GISANS-based metric. A decomposition of the effective DMI field for each slab $s$,

\begin{equation}
\label{eq:effective-dmi-field-slab}
\boldsymbol{\mathcal{W}}_s(\mathbf{q})=
\sum_{(i,j)\in s}
\mathbf{D}_{ij}\sin(\mathbf{q}\cdot\mathbf{r}_{ij}),
\end{equation}

\noindent shows that the model of Eq. \ref{eq:total-energy-spiral-ansatz} results in a finite odd component under the reflection transformation $z\rightarrow n_z-z$, giving a measurable signature of the top-bottom asymmetry of the DMI field in the mesoscopic structure and, therefore, a finite effect of the compositional gradient on this effective DMI field (see Supplementary Note 6). Finally, as an additional byproduct of the \texttt{SAGNN} predictions, we can also compute the average spin magnetic moments and compare them with those measured for sample ii.FePd (see Table~\ref{table1}). For Fe, we obtain $\sim2.78$~$\mu_B$/atom, and for Pd $\sim0.27$~$\mu_B$/atom. These values are notably smaller than the moments of fully ordered bulk Fe$_{50}$Pd$_{50}$, and together give a total moment in excellent agreement with the experimental value, of $3.11 \pm0.03$~$\mu_B$/f.u.

\subsubsection{Nature of the disorder-induced DMI: insights from the DFT dataset}

Apart from serving as a reference for training and comparison with \texttt{SAGNN} predictions on the validation set, the DFT results that constitute the database can also provide some interesting and non-trivial insights into the nature of the DMI induced by chemical disorder, particularly for Fe-Fe pairs. In Fig.~\ref{fig:parity-plots-main}(i), we show the radial dependence of the average DMI magnitude, $\overline{|\mathbf{D}_{ij}|}(r)$, computed for Fe-Fe interactions and averaged over all the 100 disordered FePd supercells (see Methods, Section~\ref{sec:methods}). In contrast to pristine FePd -- where the Fe-Fe DMI vanishes due to symmetry (Fig.~\ref{fig:FePd}) -- chemical disorder gives rise to a statistically coherent DMI field with a well-defined radial structure. The average magnitude decays approximately as a power-law, $C r^{-\alpha_{\text{eff}}}$, where $C$ is a constant and the effective exponent $\alpha_{\text{eff}} \sim 2.75$. This decay is notably slower than the canonical $r^{-3}$ behavior characteristic of RKKY-like interactions~\cite{Pajda2001,Fert1980}. This, in light of the perturbation theory developed by Fert and Levy \cite{Fert1980}, is consistent with disorder activating additional anisotropic contributions beyond a simple long-distance RKKY-like scaling. The root-mean-square DMI, $D_{\text{RMS}}$, decays even more slowly, with $\alpha_{\text{eff}} \sim 2.6$, indicating a broad distribution that includes relatively strong long-range coupling events. The DFT data further shows that short-range Fe-Fe DMI is enhanced in local environments with intermediate Pd coordination, with the largest values occurring for Fe atoms surrounded by approximately $4$-$5$ Pd nearest neighbors (in the L1$_0$ structure, the maximum coordination number in the first shell is 8 when $c \neq a$). Although the vectorial average of the DMI mostly cancels due to disorder, a finite residual component remains. Explicitly, we find a maximum ratio

\begin{equation}
\frac{|\langle \mathbf{D}_{ij} \rangle|}{D_{\text{RMS}}}
=
\frac{
\left|\sum_{(i,j)\in\mathcal{E}_k}\mathbf{D}_{ij}\right|
}{
\sqrt{
N_k \sum_{(i,j)\in\mathcal{E}_k}\left|\mathbf{D}_{ij}\right|^2
}
}
\sim 0.12,
\end{equation}

\noindent where $\mathcal{E}_k$ denotes the set of $N_k$ pairs $(i,j)$ in each $k$-th shell, corresponding to a mean DMI magnitude of about $12\%$ of the RMS value. This indicates a weak but non-negligible preferential orientation of the DMI vectors, most pronounced at intermediate distances, with a peak around $r \sim 2a$. Although modest at the pairwise level, such a residual orientation bias can become relevant collectively, because it provides a microscopic route to the net handedness detected experimentally by PA-GISANS. 

\section{Summary and Outlook}
%Now already included:
%\textcolor{red}{Genearl Note Annika: This text feels more like an outlook than a conclusion. In general We have "concluded"/interpreted already all observations withitn each section seperately. I would name this section "Summary and Outlook", then 1. give a short summary of all observations from each section, then 2. go on with the below outlook. 3. One could add an outlook on calculations in the superconducting state to examine the chirality of S/F heterostructures? If you agree to this, please state it here, and I will summarize all experimental results first}\\

%\textcolor{purple}{Helena in purple: Is it really necessary to go to this description of what we are not doing? I mean the discussion around equation 6 and following. Also, what we "anticipate"....shall we go through this? It seems to me not a good strategy to discuss what we have NOT done and what we plan to do, but I leave to Ivan's taste} 

In summary, we combined structural and magnetic characterization with first-principles-based multiscale modeling to elucidate the origin of chirality in FePd-based S/F hybrids. The experiments reveal pronounced structural disorder, including atomic intermixing and a depth-dependent defect gradient, together with finite magnetic net chirality detected by PA-GISANS. From a theoretical perspective, our results demonstrate that chemical disorder in FePd, particularly when combined with a compositional gradient, provides a realistic microscopic mechanism for the emergence of finite DMI and chiral finite-$\mathbf{q}$ spin textures.  The resulting in-plane modulation length approaches the experimentally observed range. This interpretation is corroborated by the microscopic insights provided by the DFT dataset, which reveal that disorder-induced DMI, although strongly fluctuating at the pair level, exhibits a slower decay than the canonical $r^{-3}$ behavior of RKKY-like interactions and retains a weak but finite orientation bias, most pronounced at intermediate distances, with a peak around distances $r\sim 2a$.

%\textcolor{blue}{\st{The \texttt{SAGNN} framework is essential in this context, as it preserves the local environment variability -- suppressed by shell-averaged models, such as the chosen baseline model -- and enables explicit mesoscopic modeling beyond the reach of direct DFT. Taken together, these findings support the interpretation that the chirality observed in the samples i.Nb/FePd and ii.FePd does not need to arise exclusively from interface effects, but can already emerge from chemically disordered FePd itself. }}
%%%%%%%%%%%
A central outcome of this work is the identification of chemical disorder and compositional gradients as intrinsic sources of global (net) magnetic chirality in nominally centrosymmetric magnetic alloys. In particular, our results show that the chirality observed in both i.Nb/FePd and ii.FePd does not need to originate exclusively from interface effects, but can emerge directly from the bulk FePd layer when realistic disorder is taken into account. These results are crucially enabled by the \texttt{SAGNN} framework, which allows the explicit treatment of local-environment variability of magnetic interactions. In contrast to shell-averaged or effective-medium descriptions, which suppress intra-shell fluctuations, \texttt{SAGNN} preserves the dependence of $J_{ij}$ and $\mathbf{D}_{ij}$ on the local atomic configuration, enabling realistic atomistic modeling at mesoscopic scales beyond the reach of direct DFT. This capability is essential for capturing phenomena such as magnetic frustration and emergent chirality. More broadly, this work establishes a microscopic and multiscale framework for investigating chiral spin structures in disordered or partially ordered magnetic alloys, and provides a basis for future studies that explicitly incorporate superconducting effects in S/F hybrid systems \cite{Russmann2022,Reho2024}.

The present mesoscopic model should be viewed as a controlled sufficiency test for disorder-induced chirality rather than a full reconstruction of the experimental microstructure. A natural extension of the gradient model would be to include an explicit depth-dependent long-range order parameter, $S(z)$, in order to describe more quantitatively the ordering profile observed across the FePd films.

\section{Acknowledgments}

%\IM{[Please, everyone: write your acknowledgments here.]}\\
A.S. acknowledges fruitful discussions about polarization analysis and spin transport with Wai-Tung Lee, European Spallation Source (Lund, Sweden), and acknowledges funding from the Tillväxtverket grant for parts of this work. A.S. is also very thankful for discussions and project support from Thomas Brückel, JCNS-2, Forschungszentrum Jülich GmbH, Germany. J.G.C.P., A.B.K., and H.M.P. acknowledge financial support of the Brazilian agencies CAPES, CNPq, and FAPESP (Project No 2022/10095-8). A.B.K.
acknowledges support from the INCT of Materials Informatics, and 
Spintronics and Advanced Magnetic Nanostructures (INCT-
SpinNanoMag), CNPq, Brazil. E.B. acknowledges the Swedish Research Council (grant no. 2021-06157).  I.P.M. acknowledges the Crafoord Foundation (grant No. 20231063). The simulations were enabled by resources provided by the National Laboratory for Scientific Computing (LNCC/MCTI, Brazil), CCAD-AI-UFPA (Brazil), and the National Academic Infrastructure for Supercomputing in Sweden (NAISS) at NSC Centre partially funded by the Swedish Research Council through grant agreement no. 2022-06725. 

\section{Methods}
%\textcolor{red}{Note Annika: The "Methods" section is not really different to what typically is described in the supplementary? I think we should move these sections to the supplementary (for better overview, and no need to have both}\\
\label{sec:methods}
\subsection{Experimental details}\label{sec:Suppl-Experimental}

Partially ordered L$1_0$ FePd films with high thickness have been grown using Molecular Beam Epitaxy (MBE) from DCA Instruments Finland under ultra-high vacuum and at elevated substrate temperatures $>\,$600$\,$K in codeposition of Fe and Pd \cite{Stellhorn2019}. MBE enables the growth of structural high quality thin films with monolayer precision through low growth rates. To ensure an epitaxial growth of the magnetic FePd layer, first a Cr seed layer with a following thick Pd buffer layer are deposited below FePd. A thin capping layer of Pd has been grown to prevent oxidation. In the case of Nb/FePd, a layer of Nb is deposited to study the S/F interactions. Rutherford-Backscattering (RBS) measurements reveal a composition of 1:1 for Fe and Pd atoms within the FePd layer. X-ray Reflectivity (XRR) has been used to evaluate in detail the resulting layer thicknesses of both films, which are named in the main text "i.Nb/FePd" for the S/F heterostructure and "ii.FePd", for the pure FePd layer without neighboring Nb:

\begin{itemize}
    \item \textbf{i.Nb/FePd}: Pd-cap (2\,nm)/ Nb (39\,nm)/FePd (44\,nm)/ Pd-buffer (60\,nm)/Cr (1\,nm)\\
    %SN622
    \item \textbf{ii.FePd}: Pd-cap (3\,nm)/FePd (55\,nm)/Pd-buffer (62\,nm)/Cr (2\,nm).
    %SN505
\end{itemize}

For the High-Angle Annular Dark-Field Scanning Transmission Electron Microscopy (HAADF-STEM) 
%\textcolor{green}{Juan: Is it okay to first mention this measurement with the acronyms directly? \textcolor{red}{You are right, better not}}
measurements, we have used a FEI Titan G2 80-200 ChemiSTEM at ER-C, research center Juelich \cite{ERC-Juelich}. XRD measurements have been performed using a Bruker AXS D8 Advanced system, AFM/MFM measurements have been done in the ac intermittent tapping mode of an Agilent 5400 AFM/SPM probe with HQ:NSC36/Co-Cr/Al type multi-cantilevers, and bulk magnetization for magnetic hysteresis loops have been measured using a Quantum Design Magnetic Properties Measurement System (MPMS) based on a rf-SQUID.

Bulk magnetization measurements along the out-of-plane and in-plane thin film directions yield the strength of uniaxial magnetocrystalline anisotropy, $K_u$, and the shape anisotropy, $K_{sh}$, respectively, via the integration of $M(H)$ \cite{Hubert2008}.

The structural long-range order parameter $S$ is calculated from the integrated intensities of the FePd(001) and FePd(002) XRD reflections, and quantifies the degree of chemical long-range order, ranging from the fully disordered A1/fcc limit to the fully ordered tetragonal L$1_0$ structure \cite{DoktorarbeitGehanno,Warren}.
%and relates to a combination of the disordered fcc phase and the tetragonally ordered L1$_0$-phase \\

\begin{figure}[H]
    \centering
    \includegraphics[width=1\linewidth]{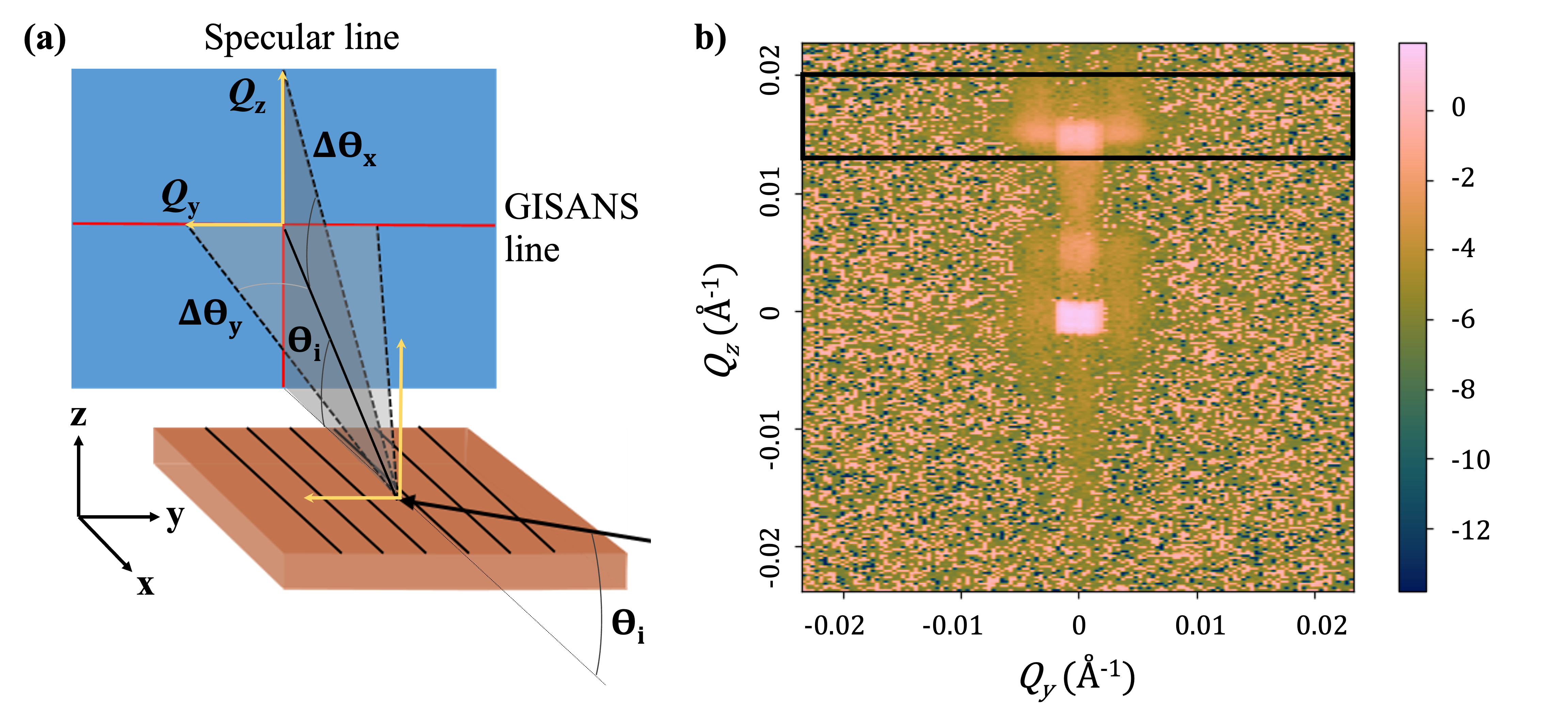}
    \caption{\textbf{GISANS setup and $Q_y$-$Q_z$ map.} (a) Schematic measurement setup of GISANS with the scattering vectors $Q_z$ and $Q_y$ along the out-of-plane $z$ and the in-plane $y$-direction, respectively.  (b) $Q_y$-$Q_z$ map of the PA-GISANS $I^{-+}$ channel of i.Nb/FePd, with $\mathbf{B}\perp[001]$. The black box indicates the region containing the GISANS and specular peaks.}
    \label{fig:2D-GISANS}
\end{figure}

Polarization-analyzed Grazing-Incidence-Small-Angle-Scattering (PA-GISANS) experiments have been performed at the D33 instrument at Institut-Laue-Langevin in Grenoble, France, under the proposal number 5-51-594. The schematic scattering geometry is shown in Fig. \ref{fig:2D-GISANS}(a).  In this schematic, the domain walls are drawn as straight lines for simplicity, whereas the actual domain-wall network in i.Nb/FePd and ii.FePd forms a maze-like pattern.
%Note that in this image, magnetic domain walls (black lines within the sample surface), leading to scattering peaks along $Q_y$, are schematically drawn as straight lines, whereas the magnetic domains of i.Nb/FePd and ii.FePd are of maze-type.
All experiments have been performed at room temperature, under a guide field of 10 G, with a wavelength of 8$\,\textnormal{\AA}$, a sample-to-detector distance of 10.7 m, and a wavelength spread of 10$\%$. For the polarization analysis, we have used the D33 supermirror V-polarizer, an RF-Flipper and a $^{3}$He-analyzer with a total flipping ratio $>$12. All measurements have been corrected for the polarization inefficiency of the polarizer, flipper and analyzer through direct beam measurements \cite{Wildes1999, Lee2023}. An example 2D scattering cross section of i.Nb/FePd is displayed in Fig. \ref{fig:2D-GISANS}(b). For an evaluation of the 2D GISANS cross-sections, we integrated the scattering intensity as function of $Q_y$ over a range of $Q_z=[0.0137,0.02]\,$nm$^{-1}$, to improve statistics (see box-inset in Fig. \ref{fig:2D-GISANS}).

\subsection{Density functional theory}
\label{sec:dft}

%\subsubsection{Density Functional Theory}

For the electronic structure calculations, we used the first-principles method RS-LMTO-ASA\cite{sonia2, RS-LMTO-ASA1}.
%We intended to study the electronic structure and magnetic properties of some systems with inversion symmetry breaking, as in the case of the presence of intermixings. For this purpose, the literature has already demonstrated that the RS-LMTO-ASA method is quite suitable \cite{Pamela}, since the eigenvalue problem within the Schr\"ondinger equation is solved directly in real space (in clusters containing on the order of $10^4$ atoms), using a tight-binding basis, and with the aid of the Haydock recursive method and the Beer-Pettifor terminator \cite{Haydock,Beer-Pettifor}.
The LMTO-ASA \cite{Andersen1975} is a linear method that gives precise results around a given energy $E_{\nu}$, generally taken as the center of the $s$, $p$, and $d$ bands. Here, the calculation considered an expression accurate to $(E-E_{\nu})$ starting from the orthogonal representation \cite{Frota-pessoa2000}. Since we are interested here in investigating the magnetic states and interactions in chemically disordered systems, an approach that solves the eigenvalue problem directly in real space is a naturally suitable choice.
In this way, the RS-LMTO-ASA scheme uses Haydock's recursion method \cite{Haydock} together with the Beer-Pettifor terminator \cite{Beer-Pettifor} to obtain the local density of states (LDOS). 

Here, the FePd bulk systems (pristine and with chemical disorder) were modeled by a big cluster containing $\sim$ 24,000 atoms located in the perfect crystal position of an fcc lattice; structure optimization was not applied as discussed in the Supplementary Note 5. In all calculations, the continued fractions have been truncated after $LL = 31$ recursion levels. After the self-consistent procedure for the collinear configuration, the magnetic interactions, namely the isotropic exchange parameters ($J_{ij}$), and the DMI vectors ($\mathbf{D}_{ij}$), are calculated using the Liechtenstein-Katsnelson-Antropov-Gubanov (LKAG) formula and extensions \cite{Szilva2023}, as implemented in RS-LMTO-ASA \cite{Frota-pessoa2000,Cardias2020}.

The spin-orbit coupling is included as a $\mathbf{L}\cdot\mathbf{S}$ term \cite{Andersen1975} computed in each variational step \cite{Frota-pessoa2004}, where $\mathbf{L}$ and $\mathbf{S}$ here denote the orbital and spin angular momentum operators, respectively. All calculations were performed within the local spin density approximation (LSDA) as parametrized by von Barth and Hedin \cite{Barth1972}. As we focus on spin-related quantities, no orbital polarization \cite{Eriksson1990} was included. 

Since the main electronic-structure code employed in this work relies on the atomic sphere approximation (ASA), which can, in principle, affect the reliability of quantities that are highly sensitive to the electronic potential, such as the magnetocrystalline anisotropy energy (MAE) \cite{Larson2003}, we opted to compute the MAE within the plane-wave pseudopotential framework as implemented in \texttt{Quantum ESPRESSO} \cite{Giannozzi2009}. The force theorem \cite{Li2014} was used to evaluate the MAE resolved over the different atomic species (Fe and Pd). All calculations were performed self-consistently, including spin-orbit coupling via fully relativistic ultrasoft pseudopotentials, with the exchange-correlation potential described by the PBEsol parametrization of the generalized gradient approximation (GGA) \cite{Perdew2008}. A $50\times50\times50$ Monkhorst-Pack $\mathbf{k}$-point grid was employed, together with kinetic-energy cutoffs of 100~Ry for the wave functions and 800~Ry for the charge density. From these calculations, we obtained a MAE of $\sim0.9$~MJ/m$^3$ for ordered L$1_0$ FePd bulk, in reasonable agreement with experiment ($\sim1.5$~MJ/m$^3$ \cite{Gehanno1997}), with $\sim84\%$ of the anisotropy attributed to the Fe sites.

%%%% REMOVIDO
%The RS-LMTO-ASA method is of great interest due to its ability to work with both exchange interaction and DMI, which, as previously mentioned, are of great importance for the formation of non-collinear spin structures such as DW, especially in systems with the presence of intermixing. 

%\subsubsection{Atomistic Spin Dynamics}
\subsection{Simulated annealing and atomistic spin dynamics}
\label{sec:asd}

%For analysis involving magnetic properties, ground state (GS), as well as micromagnetic parameter calculations,
To determine the low-energy magnetic states of our systems (Sections \ref{sec:a-localized-perspective} and \ref{sec:a-localized-perspective}), to extract critical parameters such as the Curie temperature, and to assess their robustness against thermal fluctuations, we employed simulated annealing (SA) and atomistic spin dynamics based on the stochastic Landau-Lifshitz-Gilbert (sLLG) equation, respectively, as implemented in the Uppsala Atomistic Spin Dynamics (UppASD) code \cite{Etz_2015, Skubic_2008}.

In SA, a metaheuristic method to approximate the global minimum of a function (here, the spin Hamiltonian), the spin system is heated to a temperature above the critical temperature $T_C$ and then gradually cooled down to $T \rightarrow 0$ K. Each temperature step in our simulations consisted of about $10^5$ Monte Carlo updates.

The robustness of the magnetic states against thermal fluctuations was investigated by solving the sLLG equation at a given electron bath temperature $T$, which reads

\begin{equation}
\frac{d \mathbf{m}_i}{dt} = -\gamma_L \mathbf{m}_i\times (\mathbf{B}_i+\mathbf{B}_i^{\text{fl}})-\gamma_L\frac{\alpha}{m_i}\mathbf{m}_i\times [\mathbf{m}_i \times (\mathbf{B}_i+\mathbf{B}_i^{\text{fl}})],
\label{eq:LLG}
\end{equation}

\noindent where $\gamma_L= \gamma/(1+\alpha^2)$ is the effective gyromagnetic ratio, $\alpha$ is the Gilbert damping parameter, $\mathbf{B}_i$ is the internal magnetic field at site $i$, and $\mathbf{B}_i^{\text{fl}}$ is the stochastic field that incorporates thermal effects. The internal field is derived from the spin Hamiltonian $\mathcal{H}$ as

\begin{equation}
\label{eq:internal-field}
\mathbf{B}_i = -\frac{\partial \mathcal{H}}{\partial \mathbf{m}_i},
\end{equation}

\noindent with the Hamiltonian, parameterized by electronic-structure calculations (Section \ref{sec:dft}), given by

\begin{equation}
\begin{gathered}
H = -\sum_{i\neq j}J_{ij} \mathbf{m}_i\cdot\mathbf{m}_j \\
-\sum_{i\neq j} \mathbf{D}_{ij}\cdot (\mathbf{m}_i\times \mathbf{m}_j)
-\sum_{i}K_i(\mathbf{m}_i\cdot\hat{\mathbf{e}}_z)^2,
\end{gathered}
\label{eq:Hamiltonian}
\end{equation}

\noindent where $\mathbf{m}_i$ is the normalized spin moment at site $i$ with magnitude $\mu_i$, and $K_i$ is the uniaxial magnetic anisotropy constant of site $i$, obtained as described in Section \ref{sec:dft}.

The stochastic term $\mathbf{B}_i^{\text{fl}}$ is modeled as white noise with zero mean, satisfying
\begin{equation}
\label{eq:conditions-fl}
\begin{gathered}
\langle\mathbf{B}_i^{\text{fl}}\rangle=0, \\
\langle B_{i,\chi}^{\textnormal{fl}}(t) B_{j,\varsigma}^{\textnormal{fl}}(t^{\prime}) \rangle = 
\frac{2\alpha k_{B}T}{\gamma \mu}\, \delta_{ij} \delta_{\chi\varsigma} \delta(t - t^{\prime}),
\end{gathered}
\end{equation}

\noindent where $\chi,\varsigma=\{x,y,z\}$ are Cartesian components, $k_B$ is Boltzmann’s constant, $\mu$ is the saturation magnetization at $T=0$, and $\langle\cdot\rangle$ denotes an ensemble average.

Simulations for the FePd systems were performed for supercells of size $30^3$ and larger, comprising from 54,000 atoms in the pristine case (to obtain $T_C$) up to $\sim7.6\times10^{6}$ atoms in the intermixed case (see Section \ref{sec:a-localized-perspective}). Unless otherwise specified, no external magnetic field was applied. Ground-state configurations were obtained at $T=0$, while dynamical simulations employed a Gilbert damping $\alpha=10^{-2}$, consistent with experimental values for FePd alloy films \cite{Kawai2014}.

To determine the classical ground-state of the intermixed systems, we found that the use of periodic boundary conditions constrained the SA results to a fixed topological sector, with an ordering vector dependent on the simulation box size. This indicated that the system could become trapped in local minima. To overcome this limitation, we employed open boundary conditions along all three Cartesian directions. The influence of edge effects was mitigated by systematically increasing the box size until the average energy, the spin moment per site, and the ground-state ordering vectors were converged.

%ASD simulations for FePd systems considered the cluster of $\sim$ 27,000 unit cells, being able to go from $\sim$ 54,000 up to $\sim$ 864,000 atoms, in the case of supercells; periodic boundary conditions being implemented. For GS calculation, no magnetic field was applied, and temperature was considered around $T = 0.001K$.

\subsection{Deep learning}

\subsubsection{Graph neural network architecture}

Supplementary Figure 7 shows the architecture of the Species-Aware Graph Neural Network (\texttt{SAGNN}), as used in this work. The data are divided into shells ($s$) and pair types ($p$), so that the neural network training is focused on such subspaces of the whole dataset. A total of 9197 graphs were used for training, and 2300 for validation, with nodes $\mathbf{x}_i$ and edges $\mathbf{e}_{ij}$ representing atoms and interactions, as described above. In each graph $\mathcal{G}$, the feature vectors $\mathbf{x}_i$ were embedded through a learnable linear transformation with layer normalization, resulting in the initial node representations $\mathbf{h}_i^{(1)}$. Geometric message passing with local frames was applied over all node representations $\mathbf{h}_i^{(1)}$ and their edges across $L=3$ layers. In each message-passing layer, the representations were updated according to

\begin{equation}
\label{eq:message-passing1}
\mathbf{h}_i^{(\ell+1)}=\mathbf{h}_i^{(\ell)}+s\cdot\phi\left(\mathbf{h}_i^{(\ell)},\sum_{j\in\mathcal{N}(i)}\mathbf{m}_{ij}^{(\ell)}\right),
\end{equation}

\noindent where $\ell=1,\ldots,L$ is the layer index, $\mathbf{m}_{ij}^{(\ell)}$ are the messages, $j\in\mathcal{N}(i)$ denotes the set of neighboring nodes connected via edges, and $s=0.2$ is a fixed step size. The embedded edge feature is defined as

\begin{equation}
\mathbf{h}_{ij}=\vartheta(\mathbf{e}_{ij},\varphi(r_{ij})),
\end{equation}
 
\noindent where the raw edge attributes $\mathbf{e}_{ij}$ are concatenated with the radial basis expansion $\varphi(r_{ij})$ and passed through a multilayer perceptron (MLP) $\vartheta$ with Dropout and SiLU activations. Here, we extend the bold symbol notation, typically represeting 3-component vectors in Euclidean space in the main text, to real-valued matrices. The function $\psi$ is an MLP that transforms $\mathbf{h}_j^{(\ell)}$ and $\mathbf{h}_{ij}$ into the message coefficient matrix

\begin{equation}
\mathbf{C}_{ij}^{(\ell)}=\tanh(\psi(\mathbf{h}_j^{(\ell)},\mathbf{h}_{ij}))\cdot c_{\text{scale}}\in\mathbb{R}^{H\times 3},
\end{equation}

\noindent where $c_{\text{scale}}$ is a learnable global parameter. The function $\phi$ is an MLP that transforms the aggregated message $\mathbf{m}_i^{(\ell)}=\sum_{j\in\mathcal{N}(i)}\mathbf{m}_{ij}^{(\ell)}$ together with $\mathbf{h}_i^{(\ell)}$ into the updated node representation. 

The final message has not only radial dependence via $\mathbf{C}_{ij}^{(\ell)}$ but also orientational dependence by defining a right-handed orthonormal frame 

\begin{equation}
\mathbf{B}_{ij}=[\hat{\mathbf{r}}_{ij}, \hat{\mathbf{f}}_{1,ij},\hat{\mathbf{f}}_{2,ij}]\in\mathbb{R}^{3\times 3}.
\end{equation}

Here,

\begin{equation}
\hat{\mathbf{f}}_{1,ij}=\frac{\hat{\mathbf{r}}_{ij}\times\hat{\mathbf{a}}}{|\hat{\mathbf{r}}_{ij}\times\hat{\mathbf{a}}|}, 
\end{equation}

\noindent where $\hat{\mathbf{a}}$ is chosen as the canonical axis with the smallest absolute overlap with $\hat{\mathbf{r}}_{ij}$ (for example, $[001]$ or $[010]$ when $\hat{\mathbf{r}}_{ij}$ is nearly collinear with the global $\hat{\mathbf{z}}$ axis), and $\hat{\mathbf{f}}_{2,ij}=\hat{\mathbf{r}}_{ij}\times\hat{\mathbf{f}}_{1,ij}$. Thus, the edge message is defined as

\begin{equation}
\label{eq:message-definition}
\mathbf{m}_{ij}^{(\ell)}=\mathbf{C}_{ij}^{(\ell)}\mathbf{B}_{ij}\in\mathbb{R}^{H\times 3},
\end{equation}

\noindent for a hidden dimension of $H=256$. Instead of enforcing the exact E(3)-equivariance via spherical-harmonic tensor products (see, e.g., Refs. \cite{geiger2022e3nneuclideanneuralnetworks,Batzner2022}), our model uses a local orthonormal frame $\mathbf{B}_{ij}$. The network thus predicts scalar coefficients in this frame, which are mapped back to global coordinates through $\mathbf{B}_{ij}$, which yields an SO(3)-equivariant DMI vector prediction at substantially lower cost. This is particularly suitable for chemically disordered alloys, where DMI vectors are not fixed by a single global symmetry but vary strongly with the local environment.

%In this sense, the approach used here can be seen as intermediate between E(3)-invariant methods, such as those used in chemical reaction predictions~\cite{Nippa2023}, and fully equivariant approaches \cite{Satorras2021}. This is particularly convenient for chemically disordered alloys, where DMI vectors are not fixed by a single global symmetry operation but vary strongly with the local environment, while still obeying the antisymmetry $\mathbf{D}_{ij}=-\mathbf{D}_{ji}$.
 
%This is well suited to the fact that $\mathbf{D}_{ij}$ in random alloys are not constrained by a single symmetry transformation governing all interactions (apart from the index-exchange property $\mathbf{D}_{ij}=-\mathbf{D}_{ji}$), and DMI directions can vary in non-trivial ways.

Edge distances are expanded with a learned, shell-dependent radial basis with $K=32$ harmonics,

\begin{equation}
\label{eq:radial-basis}
\begin{gathered}
\varphi(r_{ij})=
\begin{bmatrix}
\varphi_0(r_{ij}) \\
\varphi_1(r_{ij}) \\
\vdots \\
\varphi_{K-1}(r_{ij})
\end{bmatrix}
\in \mathbb{R}^K, 
\end{gathered}
\end{equation}

\noindent with

\begin{equation}
\varphi_n(r_{ij})=\cos\!\left(2k_{F,n}r_{ij}\right)\,e^{-r_{ij}/\lambda}\,f_c(r_{ij}),\quad n=0,\ldots,K-1,
\end{equation}

\noindent where the cosine cutoff is defined as $f_c(r)=\frac{1}{2}\left[\cos\!\left(\pi r/r_{\text{cut}}\right)+1\right]$ for $r\leq r_{\text{cut}}$ (and zero otherwise), with $r_{\text{cut}}=4a$. The parameters $k_{F,n}$ and $\lambda$ are also learned and constrained to be positive via a softplus transformation. The form of Eq.~\ref{eq:radial-basis} is designed to encode both oscillatory and decaying behavior of the couplings, as expected for RKKY-like interactions in systems such as FePd. Using $K=32$ harmonics allows the network to represent both slowly varying (low-frequency) and rapidly oscillating (high-frequency) radial interactions.

The node head is a linear projection producing predicted values of the local magnetic moments $\mu_i$. Edge heads use the concatenated $[\mathbf{h}_{ij},\mathbf{h}_i^{(L)},\mathbf{h}_j^{(L)}]$ to output the scalar Heisenberg exchange interactions $J_{ij}$ and DMI vectors $\mathbf{D}_{ij}$. In particular, the DMI head produces four outputs: one scalar encoding the magnitude $|\mathbf{D}_{ij}|$ and three scalar that are normalized to a unit vector giving the DMI direction in Cartesian coordinates.

To facilitate convergence, the network targets normalized variables, where the ground truth $J_{ij}$, $|\mathbf{D}_{ij}|$ and $\mu_i$ values are standardized using the mean and standard deviation pre-calculated from the training dataset for every coordination shell $s$, atomic pair type $p$ and species subset.

\subsubsection{Baseline models}

The baseline model predicted $J_{ij}$ and $|\mathbf{D}_{ij}|$ using the mean value computed over the training set for each shell $s$ and pair type $p$. Formally,

\begin{equation}
\label{eq:baseline-1}
\begin{gathered}
\bar{J}_{ij}^{\text{base}}(s,p)=\frac{1}{n(\mathcal{E}_{s,p}^{\text{train}})}\sum_{(i,j)\in\mathcal{E}_{s,p}^{\text{train}}}J_{ij} \\
|\bar{\mathbf{D}}_{ij}|^{\text{base}}(s,p)=\frac{1}{n(\mathcal{E}_{s,p}^{\text{train}})}\sum_{(i,j)\in\mathcal{E}_{s,p}^{\text{train}}}|\mathbf{D}_{ij}|
\end{gathered},
\end{equation}

\noindent where $\mathcal{E}_{s,p}^{\text{train}}$ denotes the set of edges in the training data with shell $s$ and pair type $p$, and $n(\mathcal{E}_{s,p}^{\text{train}})$ is its cardinality. Predictions for the validation set used this precomputed means without further fitting. A more detailed reasoning on this choice of baseline model is provided in Supplementary Note 5.

For the $\mathbf{D}_{ij}$ directions, as also discussed in Supplementary Note~5, one of the most physically grounded non-electronic-structure baselines for metallic magnetic systems is the Fert-Levy model. This model determines the direction of $\mathbf{D}_{ij}^{\mathrm{FL}}$ purely from geometric considerations, without requiring an electronic-structure calculation \cite{Fert1980,Gradient}. In its standard form, the DMI vector between two magnetic sites $i$ and $j$ (Fe-Fe pairs in our case) mediated by heavy-metal atoms $l$ (Pd sites) is given by the geometric kernel

\begin{equation}
\label{eq:baseline-2}
\mathbf{D}_{ij}^{\mathrm{FL}}
\;\propto\;
\sum_{l \in \mathcal{N}_{r_\text{cut}}(i,j)}
\frac{
    \left( \mathbf{r}_{li} \cdot \mathbf{r}_{lj} \right)
    \left( \mathbf{r}_{li} \times \mathbf{r}_{lj} \right)
}{
    r_{ij}\, r_{li}^{3}\, r_{lj}^{3}
},
\end{equation}

\noindent where $\mathbf{r}_{li}$ and $\mathbf{r}_{lj}$ are the vectors connecting the mediator atom $l$ to the magnetic sites $i$ and $j$, and the summation is carried out over all Pd atoms within a cutoff radius $r_{\text{cut}}$ until convergence (here, $\mathcal{N}_{r_{\text{cut}}}(i,j)$ denotes the set of all Pd atoms lying within a distance $r_c$ of either $i$ or $j$). In this work we adopt a conservative cutoff of $r_c = 30a$, which ensures convergence of the geometric series for all environments considered. Because the Fert-Levy model is grounded in a well-established microscopic mechanism for the Dzyaloshinskii-Moriya interaction, and relies solely on lattice geometry, it provides a natural and computationally inexpensive baseline to benchmark the predictive accuracy of \texttt{SAGNN} regarding the direction of $\mathbf{D}_{ij}$. A comparison of the predictive performance of both models (\texttt{SAGNN} and Fert-Levy) against the DFT reference data is provided in Supplementary Note 5.

%%%% NÃO PRECISA
%%\IM{Additional baseline experiments using a global mean predictor (i.e., independent of $s$ and $p$ indices) are provided in Supplementary Section ... [incluir um gráfico com esse baseline no SM]}

\subsubsection{Computational details}

The weighted atom-centered symmetry functions (ACSFs), here modified from the original formulation of Ref. \cite{Gastegger2018} and as described in the Supplementary Note 5, were implemented in Python, employing the open-source libraries Numpy 2.0.2 (for dataset handling) and PyTorch 2.7.1/PyTorch Geometric \cite{Fey2019} 2.6.1 (for GNN handling). To overcome the computational bottleneck of calculating the many-body angular $\mathcal{G}^2$ functions across large supercells, the descriptor generation was heavily accelerated using just-in-time (JIT) compilation and multi-threading via the Numba library. A cutoff radius of $r_c=3a$, where $a$ is the FePd lattice parameter, was used. A list of the employed descriptor parameters is given in Table \ref{tab:descriptor-paramters}. This results in $P=96$ distinct parameter sets, and $D=203=(11+2\times96)$ dimensions for each feature vector $\mathbf{x}_i$.

\begin{table}[h]
    \centering
    \caption{Employed descriptor parameters to the weighted ASCFs.}
   \begin{tabular}{@{}ccc@{}}
        \toprule
        \textbf{Parameter} & \textbf{Values} & \textbf{Dimension} \\ 
        \midrule
        $\lambda$ & $-1,+1$ & dimensionless\\ 
        $\zeta$ &  $1,4,16$ & dimensionless\\
        $\eta$ & $0.01,0.1,0.5,1.0$ & $a^{-2}$ \\
        $\mu$ & $0.1,0.5,2.0,4.0$ & $a$ \\
        \bottomrule
    \end{tabular}
    \label{tab:descriptor-paramters}
\end{table}

 Parameters were optimized using \texttt{AdamW} \cite{Loshchilov2019}. We employed parameter-specific learning rates and weight decays: the radial basis parameters were trained with a learning rate of $2\times10^{-4}$ and weight decay of $10^{-4}$, while remaining base parameters used $5\times10^{-5}$ and $10^{-5}$, respectively. The composite loss function components were weighted by $w_\mu=0.5$, $w_{J_{ij}}=1.0$ and $w_{|\mathbf{D}_{ij}|}=1.5$. Table \ref{tab:key-hyperparameters} shows key hyperparameters of the \texttt{SAGNN} model.

 \begin{table}[h]
    \centering
    \caption{Relevant hyperparameters of the \texttt{SAGNN} model and training routine.}
    \label{tab:hyperparams}
    \begin{tabular}{@{}lc@{}}
        \toprule
        \textbf{Hyperparameter} & \textbf{Value} \\
        \midrule
        Radial basis size \(K\) & 32 \\
        Radial cutoff \(r_c\) [$a$] & 4.0 \\
        Hidden dimension \(H\) & 256 \\
        Message steps \(L\) & 3 \\
        Drop‑out probability \(p\) & 0.05 \\
        Update step size $s$ & 0.2 \\
        Angular‑to‑magnitude ratio \(\lambda_{\mathrm{ang}}\) & 0.3 \\
        Optimizer & \texttt{AdamW}\\
        Base LR/Weight decay & \(5\times10^{-5}\)/\(10^{-5}\) \\
        Radial LR/Weight decay & \(2\times10^{-4}\)/\(10^{-4}\) \\
    \bottomrule
    \end{tabular}
    \label{tab:key-hyperparameters}
\end{table}

\subsubsection{Dataset filtering}

The dataset was constructed from \textit{ab-initio} calculations of 100 $4\times2\times2$ supercells (32 atoms each), composed of 50\% Pd and 50\% Fe. The atoms were arranged in the ideal fcc crystal structure, with Pd and Fe randomly distributed over the lattice sites to represent chemical disorder. Magnetic interactions were evaluated by considering every atom in each supercell as a reference site, and including all neighbors within a cutoff radius of 4 lattice parameters. This procedure yielded 388 interactions of each type (exchange and Dzyaloshinskii-Moriya) around each atom in each supercell. To generate the training data, duplicate magnetic interactions were removed according to symmetry rules for site exchange $i \leftrightarrow j$, namely $J_{ij}=J_{ji}$ and $\mathbf{D}_{ij}=-\mathbf{D}_{ji}$. This resulted in a dataset with a total of 3176 atoms (nodes) and $\sim2.36\times10^6$ interactions (edges).

\subsection{Flat spin spiral \textit{ansatz}: constrained minimization}

To minimize the flat spin-spiral \textit{ansatz} whose total energy $E(\mathbf{q},\hat{\mathbf{n}})$ is given by Eq.~\ref{eq:total-energy-spiral-ansatz} in the main text, we adopted the following procedure. For each wavevector $\mathbf{q}$, the rotation axis $\hat{\mathbf{n}}$ is variationally optimized by minimizing the combined DMI and anisotropy contributions, subject to the normalization constraint $|\hat{\mathbf{n}}|=1$. To this end, we define the effective $\mathbf{q}$-dependent DMI field $
\boldsymbol{\mathcal{W}}(\mathbf{q})=\sum_{\mathbf{r}}\bar{\mathbf{D}}(\mathbf{r})\sin(\mathbf{q}\cdot\mathbf{r})$\footnote{Note that $\boldsymbol{\mathcal{W}}(\mathbf{q})=\sum_s \boldsymbol{\mathcal{W}}_s(\mathbf{q})$, as defined in Eq.~\ref{eq:effective-dmi-field-slab} of the main text.} and represent the uniaxial anisotropy by the symmetric tensor $\mathbf{A}$,

\begin{equation}
\mathbf{A}=\sum_i K_i \left(\hat{\mathbf{e}}_i\otimes\hat{\mathbf{e}}_i\right),
\end{equation}

\noindent where $\hat{\mathbf{e}}_i$ is the local anisotropy axis at site $i$ (here, $\hat{\mathbf{e}}_i=\hat{\mathbf{e}}_z$; see Eq.~\ref{eq:Hamiltonian}). The minimization is then formulated through the Lagrangian $\mathcal{L}$:

\begin{equation}
\label{eq:lagrangean}
\mathcal{L}(\hat{\mathbf{n}},\omega)=-\boldsymbol{\mathcal{W}}^T\hat{\mathbf{n}}-\frac{1}{2}\hat{\mathbf{n}}^T\mathbf{A}\hat{ \mathbf{n}}+\frac{\omega}{2}\left(\hat{\mathbf{n}}^T\hat{\mathbf{n}}-1\right),
\end{equation}

\noindent where $\omega$ is a Lagrange multiplier, and $\hat{\mathbf{n}}^T\hat{\mathbf{n}}-1=0$ is the constraint. The stationary point with respect to $\hat{\mathbf{n}}$ is obtained from $\nabla_{\hat{\mathbf{n}}}\mathcal{L}=0$, which yields

\begin{equation}
\label{eq:condition-to-find-n}
(\omega\mathbbm{1}-\mathbf{A})\hat{\mathbf{n}}=\boldsymbol{\mathcal{W}}(\mathbf{q}),
\end{equation}

\noindent where $\mathbbm{1}$ is the identity matrix. By expressing $\hat{\mathbf{n}}$ as a function of $\omega$ from Eq.~\ref{eq:condition-to-find-n}, the unitary length constraint is reduced to the secular equation $g(\omega)=|\hat{\mathbf{n}}(\omega)|^2-1=0$, which is then solved using the Newton-Raphson iterative method. The global minimum of the total energy determines the optimal pair $(\mathbf{Q}, \hat{\mathbf{n}}_{\text{min}})$ that characterizes the ground-state magnetic structure, with in-plane spatial periodicity $\lambda = 2\pi/|\mathbf{Q}_{\parallel}|$.

The averaging in Eq. \ref{eq:total-energy-spiral-ansatz} of the main text should not be confused with the baseline model used to assess network performance. While the baseline depends only on coordination shell and pair type -- and is therefore insensitive to bond orientation and local environments -- the quantities $\bar{J}(\mathbf{r})$ and $\bar{\mathbf{D}}(\mathbf{r})$ retain directional information and anisotropies. Moreover, the gradient dependence can be preserved by performing the averaging within each slab (fixed $z$), i.e., by using $\bar{J}(\mathbf{r},z)$ and $\bar{\mathbf{D}}(\mathbf{r},z)$.

The effective-medium treatment adopted in Eq.~\ref{eq:total-energy-spiral-ansatz} of the main text should be viewed as a single-kernel description of the disordered alloy, in which the atomistic couplings are coarse-grained according to the bond vector $\mathbf r$, without considering explicit pair labels ($\varpi,\Lambda$). This approximation is well justified in the present case because the finite-$\mathbf q$ instability is dominated by the Fe-Fe couplings, while Fe-Pd interactions are predominantly ferromagnetic and Pd-Pd couplings are much weaker. The robustness of this approximation was explicitly verified by extending the constrained minimization to a label resolved form, in which $\bar{J}(\mathbf r)\rightarrow \bar{J}_{\varpi\Lambda}(\mathbf r)$ and $\bar{\mathbf D}(\mathbf r)\rightarrow \bar{\mathbf D}_{\varpi\Lambda}(\mathbf r)$, and the total energy becomes $E(\mathbf q,\hat{\mathbf n},\{\phi_\varpi\})$, depending also on the relative phases $\{\phi_\varpi\}$ between the spirals associated with different labels. In the present Fe/Pd test, this produced essentially the same optimal modulation vector $\mathbf{Q}$, supporting the treatment adopted in the main text.

\subsection{Code availability}

All electronic structure calculations were carried out using the \texttt{RS-LMTO-ASA} package. The Monte Carlo and spin-dynamics simulations were performed with \texttt{UppASD}. Both codes are publicly accessible at no cost.
%The implementation of the \texttt{SAGNN} model, together with the Python scripts used to generate all neural-network-based results reported in this work, is openly available in the SAGNN GitHub repository:
%\hyperlink{https://github.com/ivanpmiranda/sagnn}{https://github.com/ivanpmiranda/sagnn}.
The implementation of the \texttt{SAGNN} model, together with the Python scripts used to generate all neural network results reported in this work can be provided upon request from the corresponding author of the theoretical part (I.P.M.).

\subsection{Data availability}

All data generated in this study are publicly available in the Zenodo repository.

%%%%%%%%%%%%%%%%%%%%%%%%%%%%%%%%%%%%%%%%%%%%%%%%%%%%%%%%%%%%%%%%%%%%%%%%%%%%%
%\ack

\bibliographystyle{apsrev4-2}
\bibliography{Bibo}

\end{document}